\documentclass[review]{elsarticle}

\usepackage{hyperref}

\journal{Chemometrics and Intelligent Laboratory Systems}

\usepackage{amsmath}
\usepackage{amsfonts}
\usepackage{bm}
\usepackage{booktabs}
\usepackage[]{algorithm}
\usepackage{algpseudocode}
\usepackage{mathtools}
\usepackage{xcolor}

\newcommand{\spect}{\text{s}}

\newcommand{\hl}[1]{{#1}}

\begin{document}

\begin{frontmatter}

\title{Statistical Estimation of Mean Lorentzian Line Width in Spectra by Gaussian Processes}

\author[lut]{Erik Kuitunen}
\author[lut]{Matthew T. Moores}
\author[lut,aalto]{Teemu Härkönen\corref{mycorrespondingauthor}}
\cortext[mycorrespondingauthor]{Corresponding author}
\ead{teemu.h.harkonen@aalto.fi}

\address[lut]{Department of Computational Engineering, School of Engineering Sciences, LUT University, Lappeenranta, FI-53850, Finland}
\address[aalto]{Department of Electrical Engineering and Automation, Aalto University, Espoo, FI-02150, Finland}

\begin{abstract}
We propose a statistical approach for estimating the mean line width in spectra comprising Lorentzian, Gaussian, or Voigt line shapes.
Our approach uses Gaussian processes in two stages to jointly model a spectrum and its Fourier transform. We generate statistical samples for the mean line width by drawing realizations for the Fourier transform and its derivative using Markov chain Monte Carlo methods.
In addition to being fully automated, our method  enables well-calibrated uncertainty quantification of the mean line width estimate through Bayesian inference.
We validate our method using a simulation study and apply it to an experimental Raman spectrum of $\beta$-carotene.
\end{abstract}


\end{frontmatter}


\section{Introduction}
\label{sec:introduction}
An \hl{electromagnetic} spectrum consists of a number of individual bands at various locations, each with an intensity and line width.
Together, the positions of the bands characterize the sample, while their intensities can be used for quantitative analysis.
However, these bands often appear within close proximity of each other and can overlap, making them difficult to distinguish and hence complicating the analysis \cite{Antonov:2000}.
Several methods have been proposed to address this problem, such as band narrowing and curve fitting.
In particular, Fourier self-deconvolution \cite{ Kauppinen:91, fsdarticle, kauppAsump} uses a kernel function with a kernel width parameter to narrow spectral bands in order to better identify the individual spectral bands while conserving the areas of the bands.

\hl{The broadening of the bands in an electromagnetic spectrum arises from the interaction of photons and matter, resulting in the characteristic Lorentzian, Gaussian or Voigt line shapes \cite{Diem:2015}.}
\hl{This can be represented mathematically as the convolution between the bands and the line shape function, also known as the kernel function.}
\hl{Since convolution corresponds to multiplication in the Fourier domain, one can deconvolve the signal by dividing the Fourier transform of the measurement spectrum by the Fourier transform of the kernel.}
\hl{This is the key idea behind the Fourier self-deconvolution approach.}
\hl{The resulting Fourier coefficients can then be transformed back to the original signal domain, producing a more interpretable spectrum.}
\hl{Unfortunately, measurement noise corrupts this straightforward deconvolution process.}
\hl{The noise corruption is partly combated by truncating the deconvoluted signal in the Fourier domain.}
\hl{The truncated signal is extrapolated back to its original length, then the inverse Fourier transform is applied to yield a spectrum with narrower line shapes.}

\hl{However}, for the Fourier self-deconvolution to perform optimally, the kernel width parameter needs to be manually selected and to be close to the true width of the spectral bands.
As a more reasoned approach, a method for computing a mean Lorentzian bandwidth estimate of the individual bands has been proposed \cite{Loenz-Fonfria:08}.
In the paper, the authors studied an interesting analytical property of spectra consisting Lorentzian, Gaussian, or Voigt line shapes that the area-weighted average of the Lorentzian line width parameter can be estimated via the Fourier transform properties of the spectrum.
The property has very attractive features, it does not require any \textit{a priori} information on the number of line shapes, their exact functional form, or their parameters.
Additionally, the implementation is straight-forward with fast Fourier transforms (FFT).
However, the algorithm is hampered by the requirement to manually select points for a polynomial fit in the Fourier domain, making it impractical for processing large numbers of spectra.
Furthermore, the lack of uncertainty quantification makes it difficult to judge the reliability of the result.

In this paper, we extend the methodology \hl{introduced} in \cite{Loenz-Fonfria:08} by automating the estimation process while simultaneously providing well-calibrated estimates of the associated uncertainty \hl{and preserving the independence from any \textit{a priori} information on the line shapes}.
First we fit a Gaussian process to measured spectrum data.
We draw realizations from the Gaussian process to generate artificial synthetic spectra which are statistically similar to the measured spectrum.
Next, we compute Fourier transforms for each of the sampled realizations, yielding data to which we fit a second Gaussian process.
The second Gaussian process is used to sample realizations for the Fourier transforms and their derivatives, which are required for estimating the mean Lorentzian width of the spectral line shapes.
This sampling process is repeated until a desired number of samples have been generated.
We employ Markov chain Monte Carlo (MCMC) methods to generate 
these samples, thereby obtaining fully-Bayesian estimates of the unknown parameters of the two Gaussian processes.
Using MCMC, we also estimate 
a Bayesian posterior probability distribution for the area-weighted mean Lorentzian line width.
By propagating uncertainty in a consistent and rigorous manner, we are able to provide well-calibrated estimates of posterior intervals for the  parameters and the line width.
In addition to the immediate result of estimating the line width, the results can be used as an informative prior in subsequent Bayesian spectrum analysis, see for example \cite{Moores:2018, Harkonen:2020}. 

The remainder of this paper is structured as follows.
In Section \ref{sec:theory} we introduce the key mathematical properties of the original approach and introduce our statistical model for ultimately forming the mean line width posterior distribution.
Section \ref{sec:computationalDetails} contains computational details of model, such as specification of the prior distributions that were used and MCMC settings.
In Section \ref{sec:results} we present results of the proposed approach for three synthetic spectra consisting of Lorentzian, Gaussian, and Voigt line shapes.
In Section \ref{sec:experimental} we present results for an experimental spectrum of $\beta$-carotene.
In Section \ref{sec:conclusions}, we conclude the study, point out possible alternative options, and discuss possibilities for future extensions.
\section{Theory}
\label{sec:theory}
\subsection{Preliminaries}
\label{subsec:preliminaries}
An observed spectrum consisting of Lorentzian, Gaussian, or Voigt line shapes can be represented mathematically as
\begin{equation}
    \bm{S}_\text{N} \equiv \bm{S}(\bm\nu_\text{N}) = \bm{f}(\bm\nu_\text{N}; \bm{a}, \bm\ell, \bm\gamma, \bm\sigma) + \bm\epsilon_\text{N},
     \label{eq:measurementModel}
\end{equation}
where $\bm\nu_\text{N} \equiv (\nu_1, \dots, \nu_\text{N})^\top  \in \mathbb{R}^{N}$ are the discrete measurement locations, usually in units of wavenumbers (cm$^{-1}$); $\bm{S}_\text{N} \equiv \left( S(\nu_1), \dots, S(\nu_\text{N}) \right)^\top  \in \mathbb{R}^N$ are the measurements, often in scientific arbitrary units (a.u.); $f(\nu; \bm{a}, \bm\ell, \bm\gamma, \bm\sigma)$ is a  continuous mathematical model of the underlying spectral signature, with vectors of parameters $\bm{a}$ for the areas of the line shapes, $\bm\ell$ for their locations, $\bm\gamma$ for the Lorentzian line widths, and $\bm\sigma$ for the Gaussian line widths; and $\bm\epsilon_\text{N} \equiv (\epsilon_1, \dots, \epsilon_\text{N})^\top  \in \mathbb{R}^{N}$ are additive measurement errors, where $\epsilon_\text{n} \sim \mathcal{N}(0, \sigma^2_{\epsilon})$ is assumed to be white noise with unknown variance $\sigma^2_\epsilon$.
We follow \cite{Moores:2018, Harkonen:2020} in modelling the spectral signature as a linear combination of Lorentzian, Gaussian, or Voigt line shapes,
\begin{equation}
    f(\nu; \bm{a}, \bm{\ell}, \bm\gamma, \bm\sigma) = \sum\limits_{ m = 1}^M a_m \mathcal{K}( \nu - \ell_m; \gamma_m, \sigma_m),
    \label{eq:continuousSpectrumModel}
\end{equation}
where $ a_m \in \bm{a} > 0$, $ \ell_m \in \bm{\ell} > 0$, $ \gamma_m \in \bm\gamma \ge 0$, and $ \sigma_m \in \bm\sigma \ge 0$ are the parameters of the $M$ constituent line shapes, and where the line-shape function $\mathcal{K}( \nu; \gamma_m, \sigma_m)$ can be represented as
\begin{equation}
    \mathcal{K}( \nu; \gamma_m, \sigma_m) = \frac{1}{\pi \gamma_m}\frac{\gamma_m^2}{ \nu^2 + \gamma_m^2} * \frac{1}{\sqrt{2\pi \sigma_m^2}} \exp\left( - \frac{ \nu^2}{2\sigma_m^2} \right),
\end{equation}
where $\ast$ denotes convolution with respect to the wavenumbers $\nu$.
This is the Voigt line-shape function that corresponds to the Lorentzian line shape in the limit when $\sigma_m = 0$ and to the Gaussian line shape when $\gamma_m = 0$ \cite{Diem:2015}. 
As the Voigt line-shape function does not have a closed-form expression, we use the pseudo-Voigt approximation, which is defined as
\begin{equation}
    \widetilde{\mathcal{K}}(\nu; \gamma_m, \sigma_m) = \eta \cdot \frac{1}{\pi \gamma_m}\frac{\gamma_m^2}{ \nu^2 + \gamma_m^2} + (1 - \eta) \cdot \frac{1}{\sqrt{2\pi \sigma_m^2}} \exp\left( - \frac{ \nu^2}{2\sigma_m^2} \right),
\end{equation}
where $ \eta \in [0,1]$ is the mixing parameter, which is determined by the line-width parameters $ \gamma $ and $ \sigma $. For further details, see \cite{Kielkopf:73, OLIVERO1977233, Ida:00}.
For spectra such as above, we have the interesting property that
\begin{equation}
    \overline{\gamma} = \lim_{ \xi \searrow 0 } \widetilde{\gamma}(\xi) = \lim_{ \xi \searrow 0 } -\,\frac{1}{\pi \left| \widehat{S}( \xi ) \right| } \times \frac{\partial \left| \widehat{S}( \xi ) \right| }{\partial \xi} = \frac{\sum^M_{ m = 1 } a_m \gamma_m}{ \sum^M_{ m = 1 } a_m},
    \label{eq:meanLineWidth}
\end{equation}
where $\overline{\gamma}$ denotes the area-weighted average of the Lorentzian line widths and $\widehat{S}( \xi )$ denotes the Fourier transform of the spectrum $S(\nu)$ \cite{Loenz-Fonfria:08}.
It is notable that the property in Eq.~\eqref{eq:meanLineWidth} does not require any \textit{a priori} information on the line shape parameters, nor on the number of them.
In the following section, we formulate our Bayesian model for estimating the mean Lorentzian line width with a two-stage, Gaussian process approach.
\subsection{Statistical model}
\label{subsec:model}
We assume the spectrum in Eq.~\eqref{eq:measurementModel} to be distributed according to a Gaussian process:
\begin{equation}
    \bm{S}_\text{N} \sim \mathcal{GP}\left( \bm\mu(\bm\nu_\text{N}; \alpha),\, \Sigma(\bm\nu_\text{N}, \bm\nu_\text{N}; \bm\theta) + \sigma_\epsilon^2I_\text{N} \right),
    \label{eq:spectrumGP}
\end{equation}
where $\bm{S}_\text{N}$, $\bm\nu_\text{N}$ and $\sigma_\epsilon^2$ are as given above; $\bm\mu(\bm\nu_\text{N}; \alpha)$ is the mean function and $\Sigma(\bm\nu_\text{N}, \bm\nu_\text{N}; \bm\theta) \in \mathbb{R}^{ N \times N }$ is the symmetric, positive-definite covariance matrix of the Gaussian process, parameterized according to $\alpha$ and $\bm\theta$, respectively; and $I_\text{N}$ is the $N \times N$ identity matrix.
Each $i,j$th element of the covariance matrix $\Sigma(\bm\nu, \bm\nu; \bm\theta)$ is defined as
\begin{equation}
    \left[\Sigma( \bm\nu, \bm\nu; \bm\theta) \right]_{ij} \coloneqq k(\nu_i, \nu_j; \bm\theta),
     \label{eq:spectrumCovmatDef}
\end{equation}
where $i,j \in 1, \dots, M$. We use a squared-exponential kernel function:
\begin{equation} \label{eq:squaredExpKernel}
    k(\nu_i, \nu_j; \bm\theta) = \sigma_{\spect}^2  \exp\left\{ -\frac{1}{2}\frac{(\nu_i - \nu_j)^2}{ \varphi^2 } \right\}, 
\end{equation}
where  $\bm\theta \equiv ( \sigma_\text{s} ^2, \varphi)^\top$ is a vector of the GP signal variance and length scale parameters.
For the GP mean function, we use a simple constant mean, $\mu( \nu; \alpha) = \alpha$.

We formulate the task of estimating the model parameters $ \alpha$, $\bm\theta$, and $\sigma_\epsilon$ from the observed spectrum $\bm{S}_\text{N}$ as a statistical inference problem in terms of Bayes' theorem.
At its simplest, a Bayesian posterior distribution comprises two parts: the likelihood $ \mathcal{L}( \bm{S}_\text{N} \mid \alpha, \bm\theta, \sigma_\epsilon ) $ of the model fitting the data; and a prior distribution $ p( \alpha, \bm\theta, \sigma_\epsilon ) $ encoding \textit{a priori} known information regarding the model parameters.
More explicitly, a posterior distribution for the model parameters in Eq.~\eqref{eq:spectrumGP} can be given as 
\begin{equation}
    p( \alpha, \bm\theta, \sigma_\epsilon \mid \bm{S}_\text{N}) \propto \mathcal{L}( \bm{S}_\text{N} \mid \alpha, \bm\theta, \sigma_\epsilon ) \;p( \alpha, \bm\theta, \sigma_\epsilon ),
    \label{eq:spectrumGpPosterior}
\end{equation}
up to an unknown normalizing constant, $p(\bm{S}_\text{N})$.
The natural logarithm of the GP model likelihood is given by
\begin{equation}
\begin{split}
    \log\mathcal{L}(\bm{S}_\text{N} \mid \alpha, \bm\theta, \sigma_\epsilon ) = & -\frac{1}{2} \left( \bm{S}_\text{N} - \mu( \bm\nu; \alpha) \right) \left( \Sigma( \bm{\nu}, \bm{\nu}; \bm\theta ) + \sigma_\epsilon^2 I_\text{N} \right)^{-1} \left( \bm{S}_\text{N} - \mu( \bm\nu; \alpha) \right)^\top \\
    &- \frac{1}{2}\log \left| \Sigma( \bm{\nu}, \bm{\nu}; \bm\theta ) + \sigma_\epsilon^2 I_\text{N} \right| - \frac{N}{2}\log 2\pi,
\end{split}
     \label{eq:spectrumGpLogLikelihood}
\end{equation}
where $\left\vert \Sigma( \bm{\nu}, \bm{\nu}; \bm\theta ) + \sigma_\epsilon^2 I_\text{N} \right\vert$ is the matrix determinant of $ \Sigma( \bm{\nu}, \bm{\nu}; \bm\theta ) + \sigma_\epsilon^2 I_\text{N}$.
The posterior distribution $p( \alpha, \bm\theta, \sigma_\epsilon \mid \bm{S}_\text{N})$ is analytically intractable and thus we need to resort to numerical methods.
We generate random samples from the aforementioned posterior distribution using Markov chain Monte Carlo, in particular with the delayed-rejection adaptive Metropolis (DRAM) algorithm~\cite{Haario2006}.
We provide details of the prior distribution in Section \ref{sec:computationalDetails}.

\begin{figure}
    \centering
    \includegraphics[width = \textwidth]{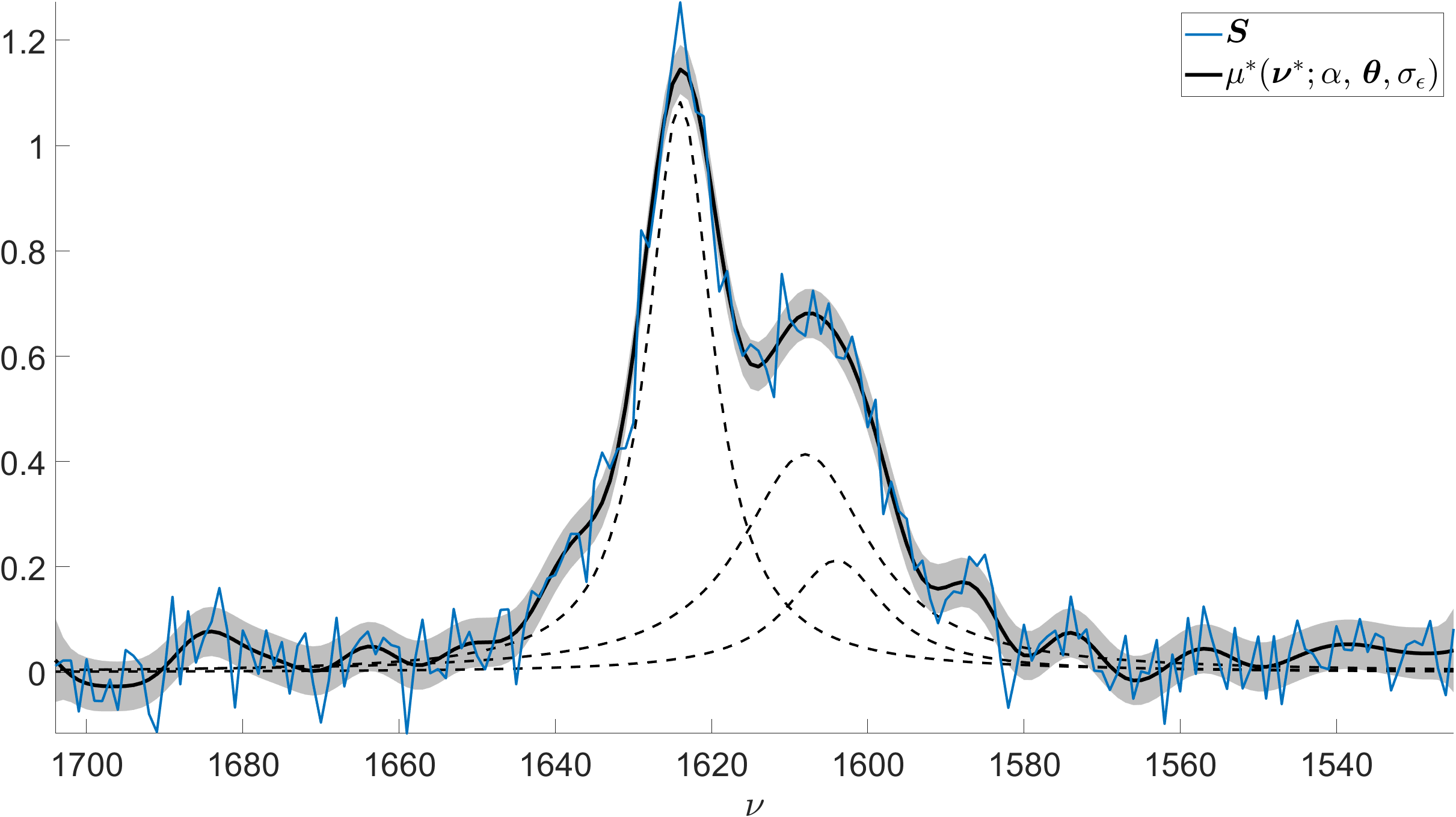}
    \includegraphics[width = \textwidth]{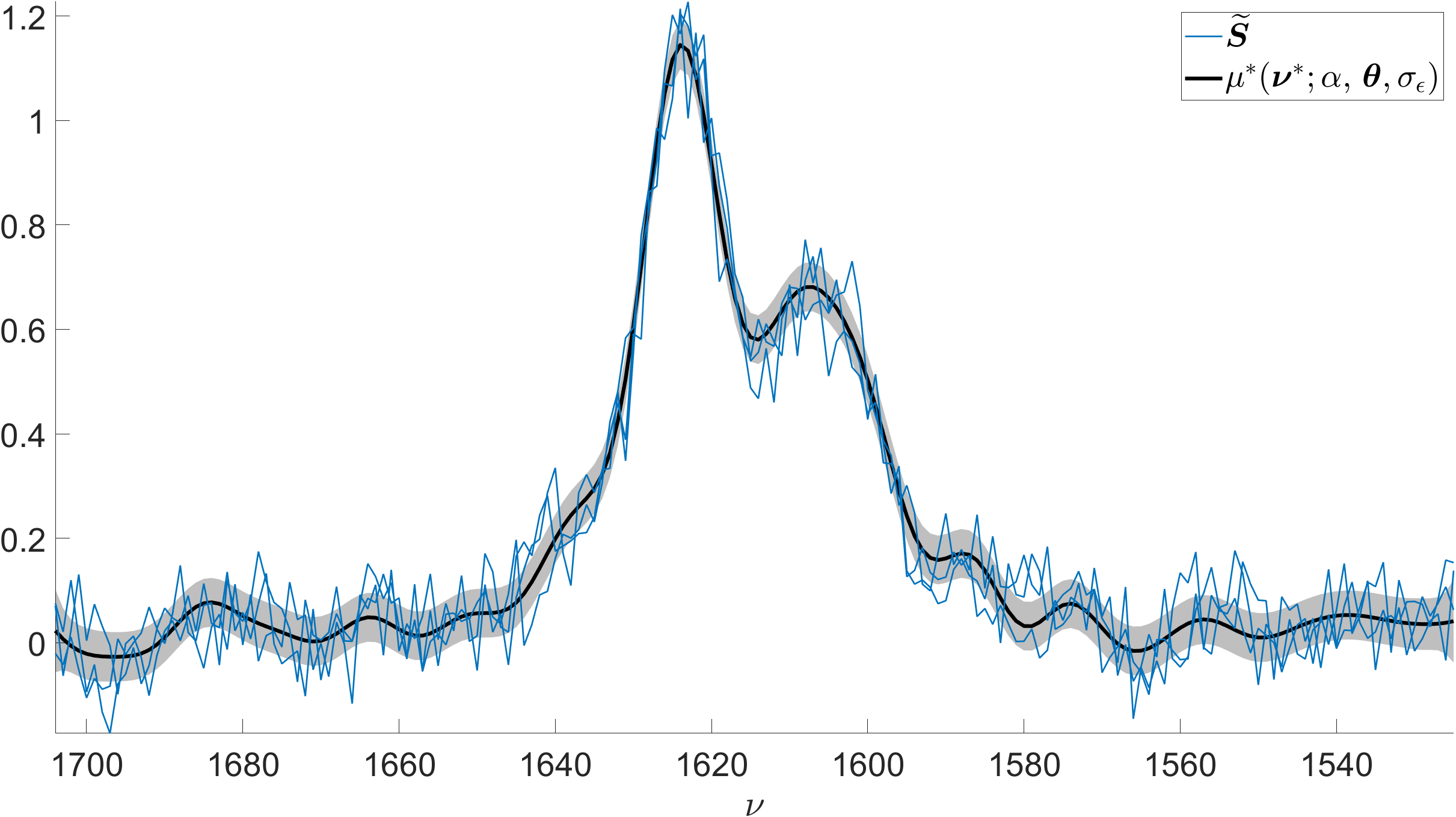}
    \caption{On top, a spectrum consisting of 3 Lorentzian line shapes in blue along with an example Gaussian process fit. The solid black line and the shaded area are the predictive mean and 95\% confidence intervals of the Gaussian process. On the bottom, realizations drawn from the Gaussian process in blue. The realizations can be used to generate additional artificial data.}
    \label{im:spectrumGpExample}
\end{figure}

Given samples from the posterior $p( \alpha, \bm\theta, \sigma_\epsilon \mid \bm{S}_\text{N})$, we can draw samples, or realizations, from the predictive distribution of the GP in Eq.~\eqref{eq:spectrumGP} using the predictive mean and covariance of the associated GP fit.
The predictive mean and covariance at prediction locations $\bm\nu^*$ are constructed with
\begin{equation} \label{eq:predMeanSpect}
    \bm\mu^*(\bm\nu^*; \alpha, \bm\theta, \sigma_\epsilon) = \Sigma( \bm\nu^*, \bm\nu; \bm\theta ) \left( \Sigma(\bm\nu, \bm\nu; \bm\theta) + \sigma_\epsilon^2 I \right)^{-1} \left(\bm{S}_\text{N} - \mu( \bm\nu; \alpha ) \right) + \mu( \bm\nu^*; \alpha),
\end{equation}
and
\begin{equation} \label{eq:predCovSpect}
    \Sigma^*( \bm\nu^*; \bm\theta, \sigma_\epsilon) = \Sigma( \bm\nu^*, \bm\nu^*; \bm\theta )- \Sigma( \bm\nu^*, \bm\nu; \bm\theta ) \left( \Sigma( \bm\nu, \bm\nu; \bm\theta) + \sigma_\epsilon^2 I \right)^{-1} \Sigma( \bm\nu^*, \bm\nu; \bm\theta)^\top.
\end{equation}
The GP is defined as a continuous random function in $\mathbb{R}$, therefore
the prediction locations $\bm\nu^*$ can be freely chosen.
We use the measurement locations as our prediction locations, $\bm\nu^* = \bm\nu_\text{N}$.
Realizations with noise $\widetilde{\bm{S}}$ can then be drawn with
\begin{equation}
    \widetilde{\bm{S}} = \bm\mu^*(\bm\nu^*; \alpha, \bm\theta, \sigma_\epsilon) + L \bm{w} + \sigma_\epsilon \bm{e},
     \label{eq:spectrumSampling}
\end{equation}
where $L$ is the lower-triangular Cholesky decomposition of the predictive covariance $\Sigma^*( \bm\nu^*; \bm\theta, \sigma_\epsilon)$, with $\bm{w} = ( w_1, \dots, w_\text{N})^\top$ and $\bm{e} = ( e_1, \dots, e_\text{N})^\top$ being vectors of independent, standard normal random variables such that $ w_n, e_n \sim \mathcal{N}(0, 1) $.
This approach can be used to generate an arbitrary amount of artificial ``measurements'' which are statistically similar to the original observed spectrum $\bm{S}_\text{N}$.
An illustration is shown in Figure \ref{im:spectrumGpExample}.
The above formulations constitute the first stage of our statistical method.

Next, we randomly select $J$ samples from the MCMC output for the posterior distribution in Eq.~\eqref{eq:spectrumGpPosterior}.
We denote these samples with $\left\{ ( \alpha, \bm\theta, \sigma_\epsilon)_1, \dots, ( \alpha, \bm\theta, \sigma_\epsilon)_J \right\}$.
For each sample $( \alpha, \bm\theta, \sigma_\epsilon)_j$, we compute the predictive mean and covariance using Eqs.~\eqref{eq:predMeanSpect} and \eqref{eq:predCovSpect} and use Eq.~\eqref{eq:spectrumSampling} to draw a realization $\widetilde{\bm{S}}_j$.
This results in a set of $J$ realizations at $N$ wavenumber locations $\bm\nu^*$.
Then we compute the FFT of each sample, which we use to model the Fourier transforms in Eq.~\eqref{eq:meanLineWidth}.
As we are only interested in the behaviour of Eq.~\eqref{eq:meanLineWidth} in the limit as $\xi$ approaches zero, we additionally truncate the FFTs of the realizations to an arbitrary length, $P$.
This also eases the computational burden by reducing the size of the data set.
We denote the data set consisting of the truncated FFTs by $\bm{Z} = \left( \widehat{\bm{S}}_1^{1:P}, \dots, \widehat{\bm{S}}_J^{1:P} \right)^\top \in \mathbb{R}^{JP}$ where $\widehat{\bm{S}}_j$ is the FFT of $\widetilde{\bm{S}}_j$.
The corresponding frequency locations are denoted by $\bm\xi = ( 0, \dots, \xi_P, \dots, 0, \dots, \xi_P)^\top \in \mathbb{R}^{JP}$ where the locations $( 0, \dots, \xi_P)^\top$ are repeated $J$ times, as the locations are identical for each $\widehat{\bm{S}}_j^{1:P}$ in $\bm{Z}$.
With the above, we are ready to formulate the statistical model for the Fourier-transformed realizations, similarly to what we did for the measurement spectrum.

We assume $\bm{Z}$ to be distributed according to a Gaussian process:
\begin{equation}
    \bm{Z} \sim \mathcal{ GP}( \phi( \bm\xi; \bm\beta), C( \bm\xi, \bm\xi; \bm\psi) + \sigma_z^2 I),
     \label{eq:fftGpModel}
\end{equation}
where $\phi( \bm\xi; \bm\beta) \in \mathbb{R}^{JP}$ denotes the mean function and $C( \bm\xi, \bm\xi; \bm\psi) \in \mathbb{R}^{ JP \times JP }$ denotes the covariance matrix, parameterized according to $\bm{\beta}$ and $\bm\psi$; and with independent and identically-distributed white noise, $\sigma_z^2 I$.
We again use the squared exponential kernel
\begin{equation}
    c_{0,0}( \xi_i, \xi_j; \bm\psi) = \sigma_c^2  \exp\left\{ -\frac{1}{2}\frac{( \xi_i - \xi_j )^2}{ \lambda^2 } \right\},
     \label{eq:squaredExpKernelFFT}
\end{equation}
where $i,j \in ( 1, \dots, JP)^T$ $\bm\psi = ( \sigma_c^2, \lambda)^T$ is a vector of the GP signal and length scale parameters.
\hl{Note that the Fourier transforms of the Lorentzian, Gaussian, or Voigt line shapes are infinitely differentiable, except at the origin.}
\hl{This implies smoothness in the truncated FFTs $\bm{Z}$, justifying the use of the smooth squared exponential kernel.}
Each element of the covariance matrix $C( \bm\xi, \bm\xi; \bm\psi)$ is constructed with the covariance function $c_{0,0}( \xi_i, \xi_j; \bm\psi)$ as in Eq.~\eqref{eq:spectrumCovmatDef}.
For the GP mean function, we now use an exponential function $\phi( \xi; \bm\beta) = \beta_0 \exp(\beta_1 \xi)$ where $\bm\beta = ( \beta_0, \beta_1)^T$.
The posterior distribution for the model parameters in Eq.~\eqref{eq:fftGpModel} is given as
\begin{equation}
    p( \bm\beta, \bm\psi, \sigma_z \mid \bm{Z}) \propto \mathcal{L}( \bm{Z} \mid \bm\beta, \bm\psi, \sigma_z ) p( \bm\beta, \bm\psi, \sigma_z ),
    \label{eq:fftGpPosterior}
\end{equation}
where $ \mathcal{L}( \bm{Z} \mid \bm\beta, \bm\psi, \sigma_z ) $ is the model likelihood and $ p( \bm\beta, \bm\psi, \sigma_z ) $ is the prior distribution for the model parameters.
The log-likelihood of the model is now given as
\begin{equation}
\begin{split}
    \log\mathcal{L}( \bm{Z} \mid \bm\beta, \bm\psi, \sigma_z ) = & -\frac{1}{2} \left( \bm{Z} - \phi( \bm\xi; \bm\beta) \right) \left( C( \bm{\xi}, \bm{\xi}; \bm\psi ) + \sigma_z^2 I \right)^{-1} \left( \bm{Z} - \phi( \bm\xi; \bm\beta) \right)^T \\
    &- \frac{1}{2}\log \left\vert \Sigma( \bm\xi, \bm\xi; \bm\psi ) + \sigma_z^2 I \right\vert - \frac{JP}{2}\log 2\pi.
\end{split}
     \label{eq:FTGpLogLikelihood}
\end{equation}
We document the prior distributions in Section \ref{sec:computationalDetails}.
To obtain samples from the posterior distribution in Eq.~\eqref{eq:fftGpPosterior}, we again employ the DRAM algorithm.

Next, we draw samples in order to model the Fourier transform and its derivative in Eq.~\eqref{eq:meanLineWidth}.
However, this requires some additional mathematical machinery, see for example \cite{Rasmussen2005, herbsusmannDerivativesGaussian}, in comparison to Eq.~\eqref{eq:spectrumSampling} which we will define below.
The predictive mean for the Fourier transform and its derivative at prediction locations $\bm\xi^*$ are now given as 
\begin{equation} \label{eq:predMeanFT}
    \begin{bmatrix}
        \phi^*(\bm{ \xi}^*; \bm\beta, \bm\psi, \sigma_z)\\
        \phi'^*(\bm{ \xi}^*; \bm\beta, \bm\psi, \sigma_z)
    \end{bmatrix}
     = C( \bm\xi^*, \bm\xi; \bm\psi) \left( C( \bm\xi, \bm\xi; \bm\psi) + \sigma_z^2 I \right)^{-1} \left( \bm{Z} - \phi( \bm\xi; \bm\beta) \right) + 
     \begin{bmatrix}
        \phi( \bm\xi^*; \bm\beta)\\
        \phi'( \bm\xi^*; \bm\beta)
     \end{bmatrix}
\end{equation}
where $\phi'( \bm\xi^*; \bm\beta)$ is the derivative of $\phi( \bm\xi^*; \bm\beta)$ with respect to $\xi$ and the corresponding predictive covariance is given as
\begin{equation} \label{eq:predCovFT}
    C^*( \bm\xi^*; \bm\psi, \sigma_z) = C( \bm\xi^*, \bm\xi^*; \bm\psi) - C( \bm\xi^*, \bm\xi; \bm\psi) \left( C( \bm\xi, \bm\xi; \bm\psi ) + \sigma^2_z I \right)^{ -1} C( \bm\xi^*, \bm\xi; \bm\psi )^T,
\end{equation}
where the covariance matrices $ C( \bm\xi^*, \bm\xi; \bm\psi) $ and $ C( \bm\xi^*, \bm\xi^*; \bm\psi) $ are constructed as
\begin{equation}
    C( \bm{ \xi}^*, \bm{\xi}; \bm{ \theta}) = 
    \begin{bmatrix}
        c_{0,0}( \bm\xi^*, \bm\xi; \bm\psi ) \\
        c_{1,0}( \bm\xi^*, \bm\xi; \bm\psi )
    \end{bmatrix},
\end{equation}
and
\begin{equation}
    C(\bm{\xi}^*, \bm{\xi}^*; \bm{\theta}) =
    \begin{bmatrix}
        c_{0,0}( \bm\xi^*, \bm\xi^*; \bm\psi ) & c_{0,1}( \bm\xi^*, \bm\xi^*; \bm\psi ) \\
        c_{1,0}( \bm\xi^*, \bm\xi^*; \bm\psi ) & c_{1,1}( \bm\xi^*, \bm\xi^*; \bm\psi )
    \end{bmatrix},
\end{equation}
where the $ij$th elements of each of the individual covariance matrices are defined as
\begin{equation}
    \left[ C_{ \bm\cdot,\bm\cdot}( \bm\xi, \bm\xi; \bm\psi) \right]_{ij} \coloneqq c_{ \bm\cdot,\bm\cdot}( \xi_i, \xi_j; \bm\psi).
     \label{eq:fftCovmatDef}
\end{equation}
The three additional covariance functions can be obtained by differentiating the covariance function $c_{0,0}( \xi_i, \xi_j; \bm\psi)$ defined in Eq.~\eqref{eq:squaredExpKernelFFT}:
\begin{equation}
    \begin{alignedat}{2}
        c_{0,1}( \xi_i, \xi_j; \bm\psi ) &= \frac{\partial c_{0,0}(\xi_i, \xi_j; \bm\psi )}{ \partial \xi_j } &&= \frac{ \sigma^2_c }{ \lambda^2 } ( \xi_i - \xi_j ) \exp\left\{ -\frac{1}{2}\frac{(\xi_i - \xi_j)^2}{ \lambda^2 } \right\}, \\
        c_{1,0}(\xi_i, \xi_j; \bm\psi ) &= \frac{\partial c_{0,0}(\xi_i, \xi_j; \bm\psi) }{ \partial \xi_i } &&= \frac{ \sigma^2_c }{ \lambda^2 }( \xi_i - \xi_j ) \exp\left\{ -\frac{1}{2} \frac{ (\xi_i - \xi_j )^2 }{ \lambda^2 } \right\}, \\
        c_{1,1}(\xi_i, \xi_j; \bm\psi ) &= \frac{\partial^2 c_{0,0}(\xi_i, \xi_j; \bm\psi ) }{ \partial\xi_j \partial\xi_i } &&= \frac{ \sigma_c^2 }{ \lambda^2 } \left( \lambda^2 - ( \xi_i - \xi_j )^2 \right) \exp\left\{ -\frac{1}{2}\frac{(\xi_i - \xi_j)^2}{ \lambda^2 } \right\},
    \end{alignedat}
    \label{eq:sqExpCovDerivatives}
\end{equation}
where $ c_{0,1}( \xi_i, \xi_j; \bm\psi ) $ along with $ c_{1,0}( \xi_i, \xi_j; \bm\psi ) $ model the covariance between a realization and its derivative and $ c_{1,1}( \xi_i, \xi_j; \bm\psi ) $ models the covariance of a derivate realization similarly to how $ c_{0,0}( \xi_i, \xi_j; \bm\psi ) $ models the covariance of a realization.

\begin{figure}
    \centering
    \includegraphics[width = 0.495\textwidth]{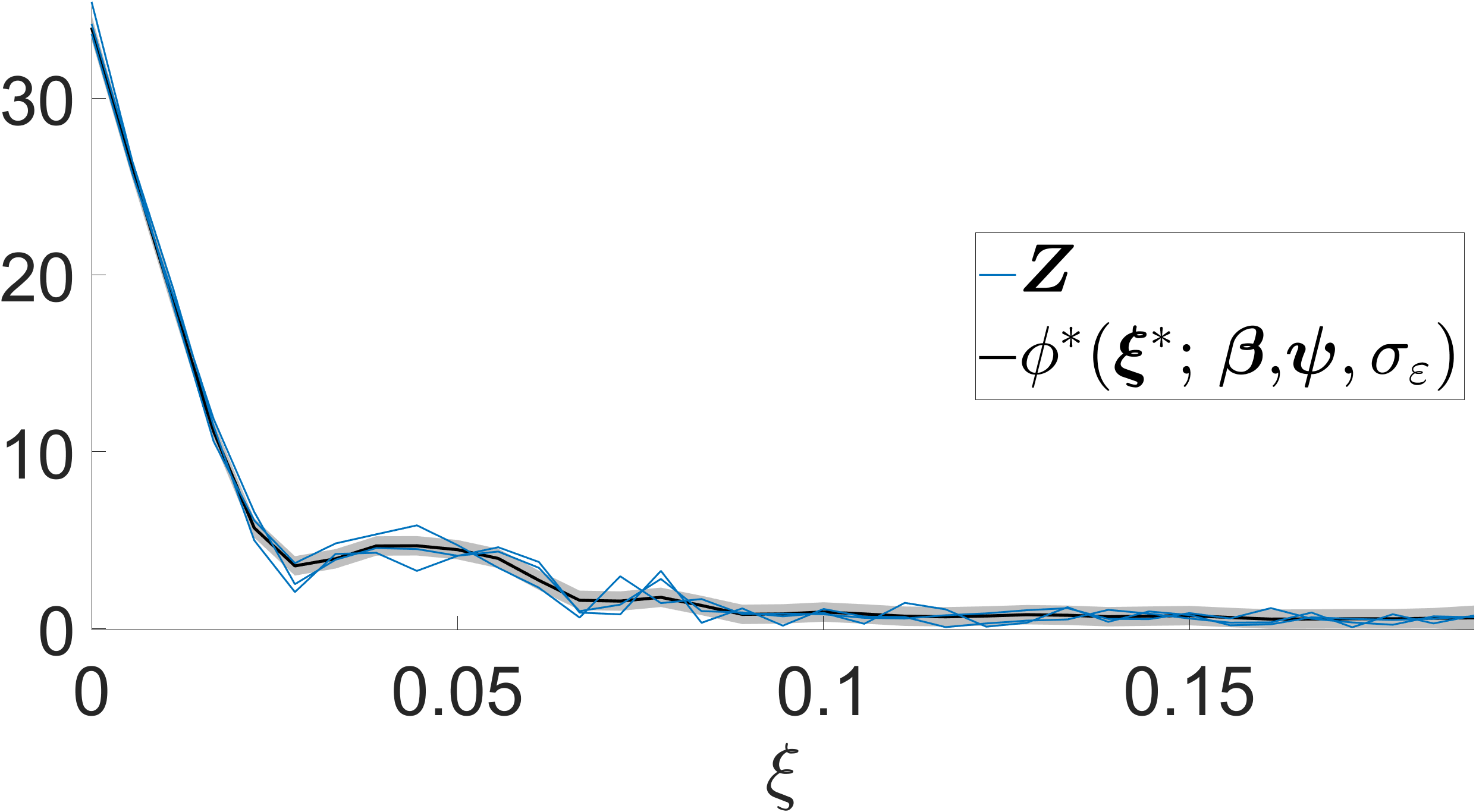}
    \includegraphics[width = 0.495\textwidth]{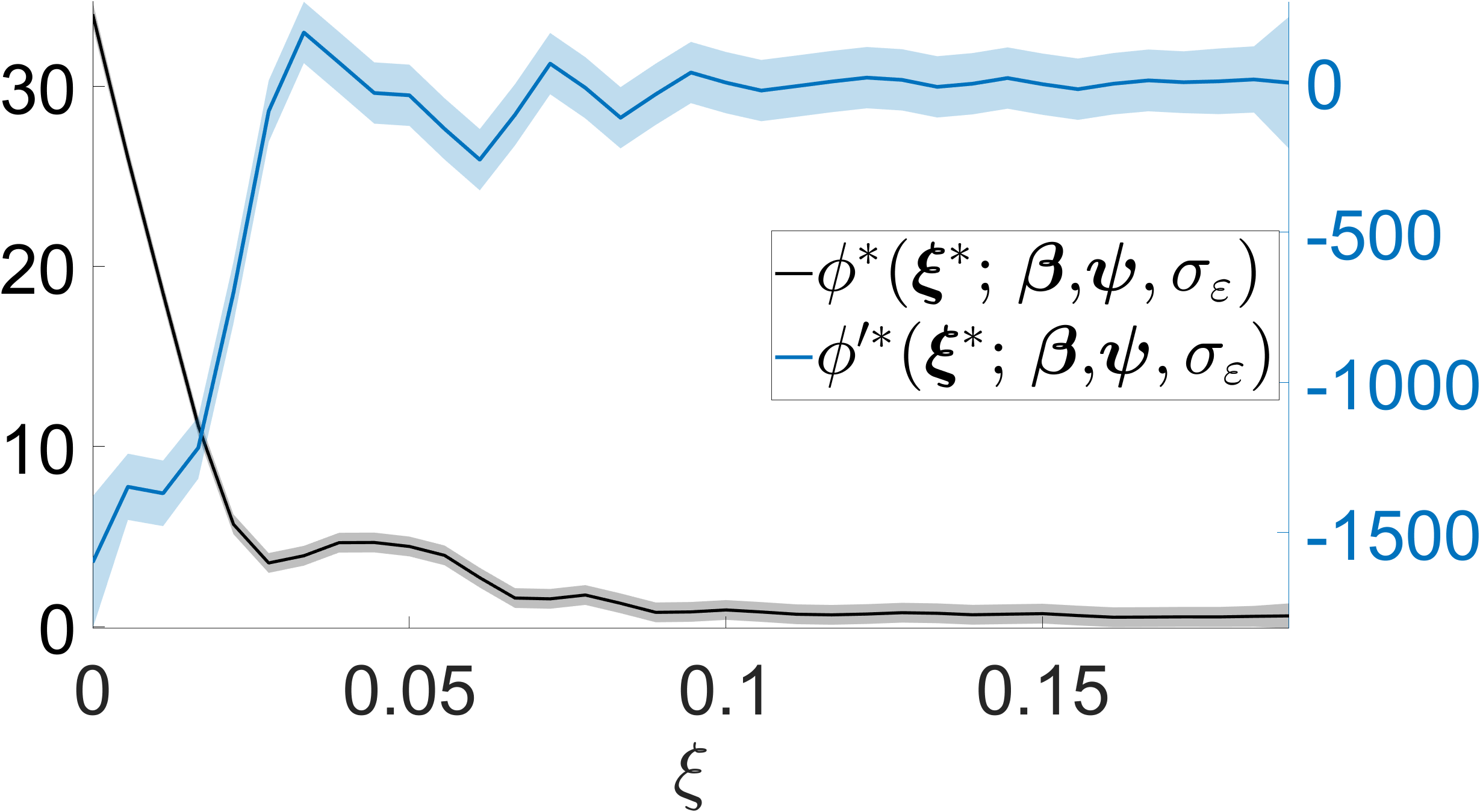}
    \includegraphics[width = 0.495\textwidth]{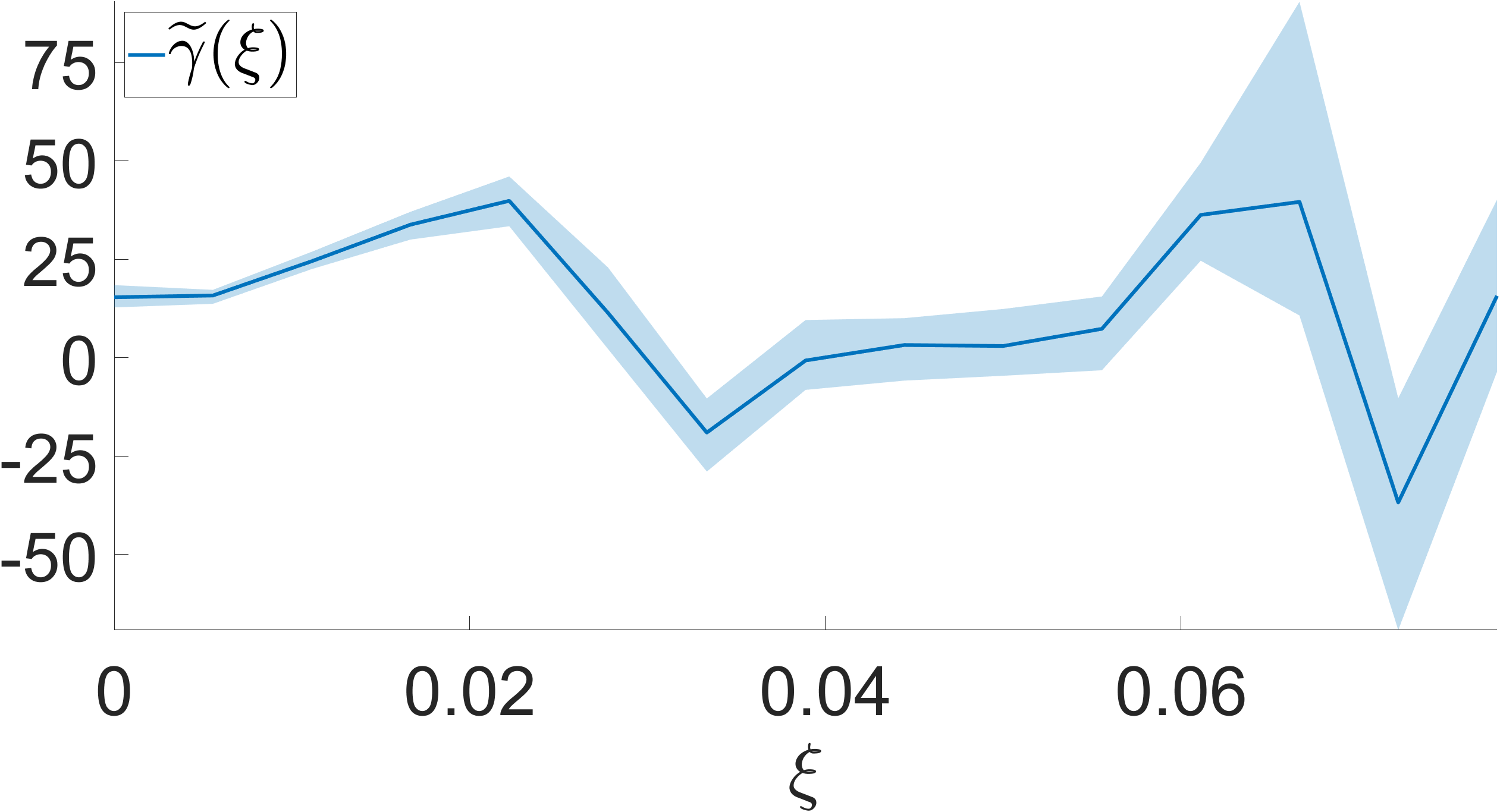}
    \includegraphics[width = 0.495\textwidth]{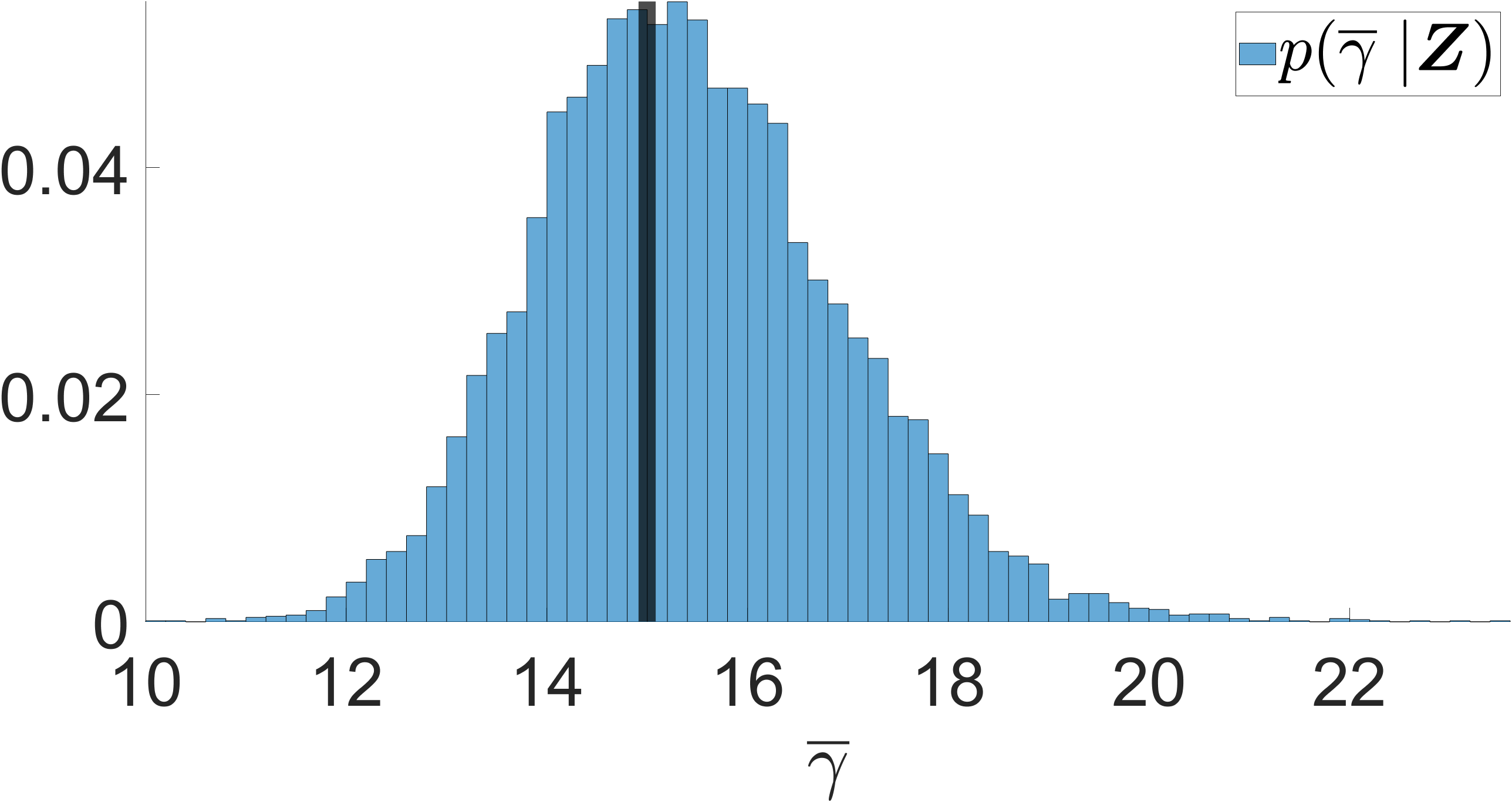}
    \caption{On the top left, truncated FFTs computed for Gaussian process realizations, in blue, shown in Figure \ref{im:spectrumGpExample} along with the associated Gaussian process fit. The solid black line corresponds to the predictive mean $\phi^*(\bm{ \xi}^*; \bm\beta, \bm\psi, \sigma_z)$ and the shaded area is the 95\% confidence interval. On top right, the same GP is shown in black and gray along with the the predictive mean for the derivative $\phi'^*(\bm{ \xi}^*; \bm\beta, \bm\psi, \sigma_z)$ and its 95\% confidence interval. On the bottom left, the expectation and 95\% confidence interval for $\widetilde{\gamma}(\nu)$ obtained by sampling Eq.~\eqref{eq:meanLineWidthEstimate}. The behaviour of $\widetilde{\gamma}(\nu)$ at $\xi = 0$ corresponds to the posterior distribution for the mean Lorentzian line width, $p( \overline{\gamma} \mid \bm{Z} )$. The true value used to generate the data in Figure \ref{im:spectrumGpExample} is illustrated with a solid vertical black line.}
    \label{im:fftGpExample}
\end{figure}

Finally, we are able to draw realizations to model the Fourier transforms in Eq.~\eqref{eq:meanLineWidth} with
\begin{equation}
    \begin{bmatrix}
        \widetilde{\bm{Z}} \\
        \widetilde{\bm{Z}}'
    \end{bmatrix}
    =
    \begin{bmatrix}
        \phi^*(\bm{ \xi}^*; \bm\beta, \bm\psi, \sigma_z)\\
        \phi'^*(\bm{ \xi}^*; \bm\beta, \bm\psi, \sigma_z)
    \end{bmatrix} + Q \bm{r},
     \label{eq:fftSampling}
\end{equation}
where $\widetilde{\bm{Z}}$ and $\widetilde{\bm{Z}}'$ denote realizations for the Fourier transform and its derivative, respectively, $Q$ is the lower triangular Cholesky decomposition matrix of the predictive covariance $ C^*( \bm\xi^*; \bm\psi, \sigma_z) $ and $\bm{r} = ( r_1, \dots, r_\text{N})^T$ is a vector of standard normal variables such that $ r_n \sim \mathcal{N}(0, 1) $.
We can then compute an estimate for the mean Lorentzian line width with
\begin{equation}
    \overline{\gamma} \sim -\frac{ \widetilde{\bm{Z}}'_0 }{ \pi \widetilde{\bm{Z}}_0 },
    \label{eq:meanLineWidthEstimate}
\end{equation}
where $ \widetilde{\bm{Z}}_0 $ and $ \widetilde{\bm{Z}}'_0 $ are the realizations evaluated at $\xi = 0$.
We sample estimates for the mean gamma line width $\overline{\gamma}$ by repeatedly sampling $( \bm\beta, \bm\psi, \sigma_z)$ from the MCMC chain for the posterior distribution in Eq.~\eqref{eq:fftGpPosterior} and by computing Eq.~\eqref{eq:meanLineWidthEstimate}.
This yields us our ultimate goal, a posterior distribution for $\overline{\gamma}$ which we denote by $ p( \overline{\gamma} \mid \bm{Z})$.
We show an illustration of $\bm{Z}$ with a respective GP fit, the respective GP predictive distribution for the derivative, mean line width function $\widetilde{\gamma}(\xi)$, and the posterior distribution $p( \overline{\gamma} \mid \bm{Z})$ in Figure \ref{im:fftGpExample}.
We summarize the above in Algorithm \ref{alg:GPWidthEstimation}.
\begin{algorithm}[H]
    \caption{Two-stage Gaussian process estimation of mean Lorentzian lime width}
    \label{alg:GPWidthEstimation}
    \begin{algorithmic}
        \State Step 1: Fit a GP to the measurement spectrum $\bm{S}$ with MCMC using Eq.~\eqref{eq:spectrumGpPosterior}.
        \State Step 2: Draw $J$ realizations $\widetilde{\bm{S}}$ via Eq.~\eqref{eq:spectrumSampling}.
        \State Step 3: Form the truncated FFT data set $\bm{Z}$ and fit a GP using Eq.~\eqref{eq:fftGpPosterior} and MCMC.
        \State Step 4: Draw $J_z$ Fourier-domain realizations with Eq.~\eqref{eq:fftSampling}.
        \State Step 5: Compute the mean Lorentzian line width for the $J_z$ realizations with Eq.~\eqref{eq:meanLineWidthEstimate}.
    \end{algorithmic}
\end{algorithm}
\section{Computational details}
\label{sec:computationalDetails}
The posterior distributions defined in Eqs.~\eqref{eq:spectrumGpPosterior} and \eqref{eq:fftGpPosterior} have prior distributions $ p( \alpha, \bm\theta, \sigma_\epsilon ) $ and $ p( \bm\beta, \bm\psi, \sigma_z ) $.
We use simple uniform priors for all our parameters to enforce positivity of the parameters which are either physically or mathematically positive such as the mean Lorentzian line width $\overline{\gamma}$ or standard deviation parameters such as $\sigma_\epsilon$ and $\sigma_z$.
\hl{The upper bounds defined by the prior distributions $\ell \sim \mathcal{U}( 0, 2\left( \nu_\text{N} - \nu_1 \right))$ and $\lambda \sim \mathcal{U}( 0, 3\xi_P)$ for the length scales correspond to cases where the measurement spectrum $\bm{S}_\text{N}$ or its FFT would be practically constant.
For the exponential mean function of the FFT derivatives $\phi( \xi; \bm\beta) = \beta_0 \exp(\beta_1 \xi)$, at $\xi = 0$ the value $ \beta_1 \sim \mathcal{U}( 0, 10 \times \max\bm{Z})$ is bounded above to be less than 10 times the maximum of the sampled FFT values $\bm{Z}$.}
Additionally, all our prior distributions are modelled as jointly independent distributions such that $ p( \alpha, \bm\theta, \sigma_\epsilon ) = p( \alpha ) p( \sigma_s ) p( \ell ) p( \sigma_\epsilon )$ and $ p( \bm\beta, \bm\psi, \sigma_z ) = p( \beta_1 ) p( \beta_2 ) p( \sigma_c ) p( \lambda ) p( \sigma_z )$.
We detail these individual distributions in Table \ref{tb:priors}.
We also enforce positivity on the posterior $ p( \overline{\gamma} \mid \bm{Z} ) $.
\begin{table}
\caption{Prior distributions for the GP parameters.}
\centering
\begin{tabular}{c c c c}
\toprule
Parameter & Distribution & Parameter & Distribution \\
\midrule
$p( \sigma_s )$ & $\mathcal{U}( 0, \infty)$ & $p( \sigma_c )$ & $\mathcal{U}( 0, \infty)$ \\
$p( \ell )$ & $\mathcal{U}( 0, 2\left( \nu_N - \nu_1 \right))$ & $p( \lambda )$ & $\mathcal{U}( 0, 3\xi_P)$ \\
$p( \sigma_\epsilon )$ & $\mathcal{U}( 0, \infty)$ & $p( \sigma_z )$ & $\mathcal{U}( 0, \infty)$ \\$p( \alpha )$ & $\mathcal{U}( 0, \infty)$ & $p( \beta_1 )$ & $\mathcal{U}( 0, 10 
\times \max\bm{Z})$ \\
 & & $p( \beta_2 )$ & $\mathcal{U}( -\infty, \infty)$ \\
\bottomrule
\end{tabular}
\label{tb:priors}
\end{table}

We run the DRAM algorithm to sample both posterior distributions $p( \alpha, \bm\theta, \sigma_\epsilon \mid \bm{S})$ and $p( \bm\beta, \bm\psi, \sigma_z \mid \bm{Z})$ in Eqs.~\eqref{eq:spectrumGpPosterior} and \eqref{eq:fftGpPosterior} with 3 proposal steps per each MCMC iteration and with a chain length of $50000$.
We use a burn-in of $25000$ \hl{and a truncation parameter $P = 30$.
In addition, we perform a sensitivity analysis for the posterior distribution $ p( \overline{\gamma} \mid \bm{Z} ) $ with respect to the truncation parameter using $P \in ( 20, 25, \dots, 95, 100)^T$ for a synthetic spectrum consisting of Lorentzian line shapes that we describe in the following Section.}
\section{Numerical examples}
\label{sec:results}
We apply the method proposed in Section \ref{sec:theory} to three synthetic spectra with additive noise.
The three synthetic data sets consist of a spectrum simulated with $ M = 8 $ Lorentzian line shapes, one simulated with $ M = 10 $ Gaussian line shapes, and lastly of a spectrum with $ M = 6 $ Voigt line shapes.
All parameters, $( \bm{a}, \bm{l}, \bm\gamma, \bm\sigma)$, and noise level of the synthetic Lorentzian and Gaussian spectra are randomly sampled from uniform distributions which are detailed in Table \ref{tb:spectrumParameters}. The synthetic Voigt spectrum is calculated using pseudo-Voigt approximation.
A log-normal prior $\delta_m \sim \mathrm{Lognormal}(\mu_{V}, \sigma_V^2)$ with $\sigma_V^2 = 0.16$ and $\mu_V = \ln(25) - \sigma_V^2/2$ is used for Voigt line width, and uniform prior $\mathcal{U}(0, \delta_m)$ is used for the Lorentzian width needed for the calculation of pseudo-Voigt approximation. The same priors for parameters $\bm{a}$ and $\bm{l}$ are used as with Lorentzian and Gaussian spectra.

\begin{figure}[H]
    \centering
    \includegraphics[width = \textwidth]{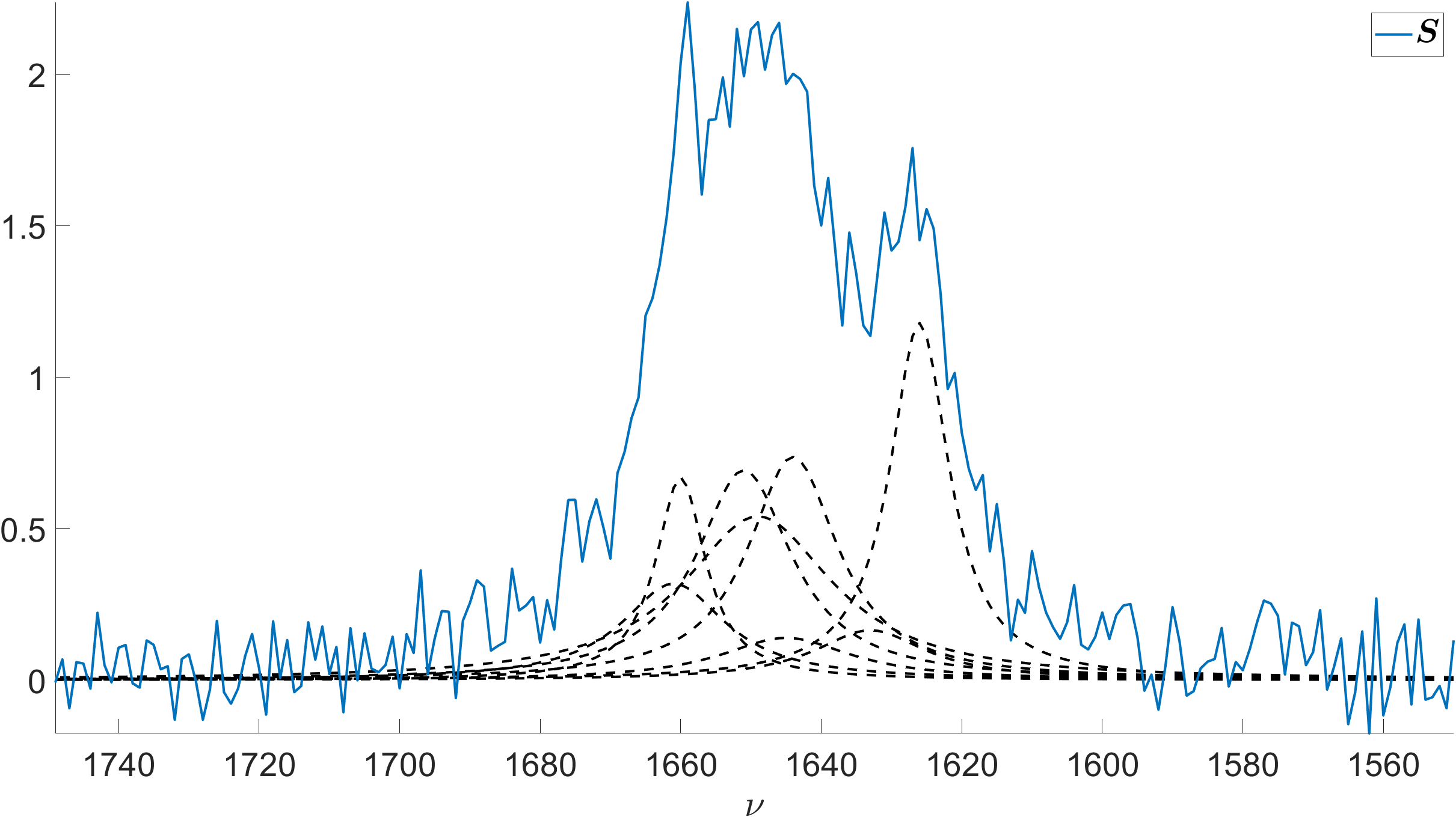}
    \includegraphics[width = \textwidth]{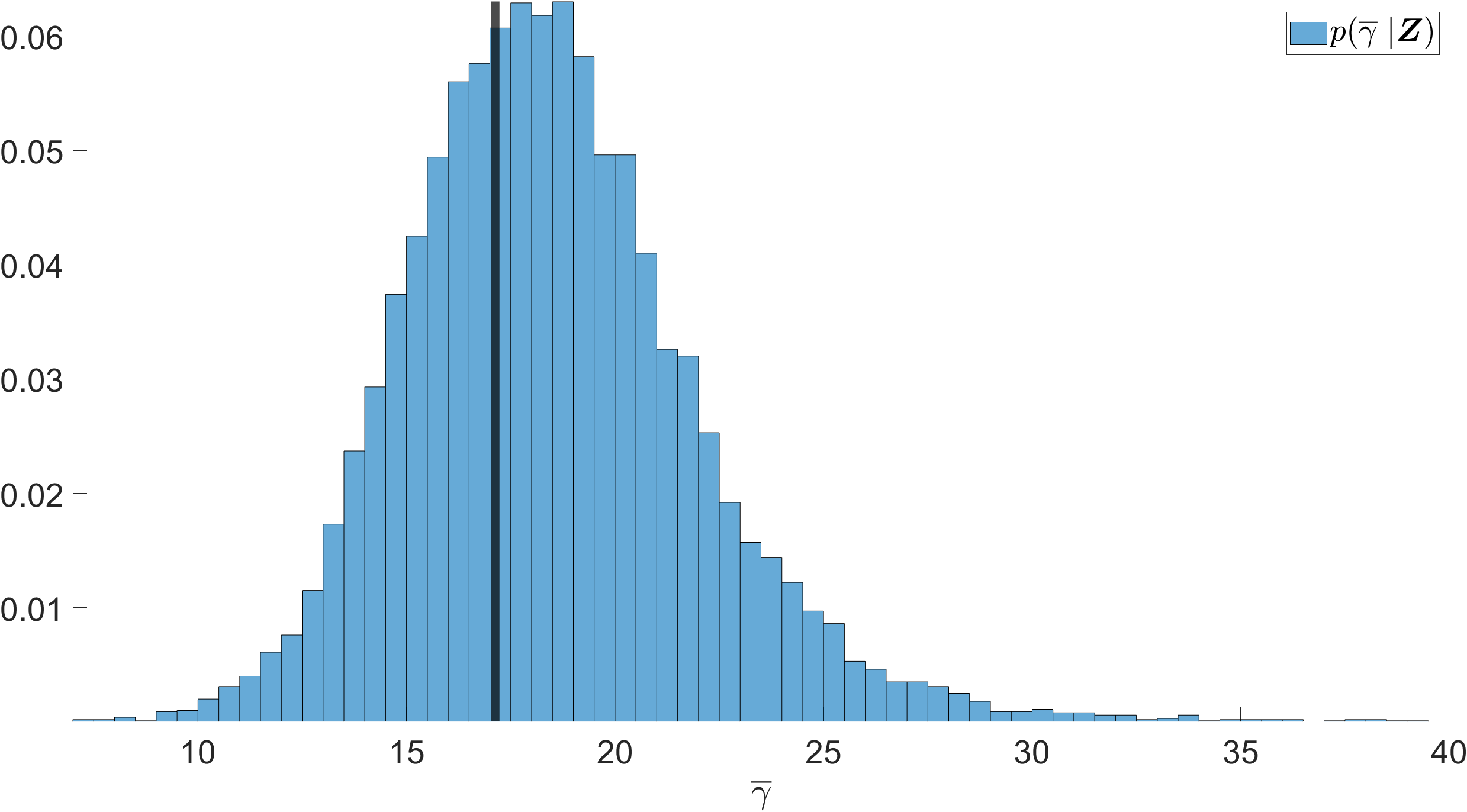}
    \caption{On top, a spectrum consisting of $ M = 8 $ Lorentzian line shapes in blue along with the individual overlapping line shapes illustrated with black dashed lines. On bottom, the obtained posterior distribution $p( \overline{\gamma} \mid \bm{Z} )$ for the area-weighted mean Lorentzian line width. The true value for $\overline{\gamma} = 17.12$ used to generate the data is shown with a vertical solid line.}
    \label{im:resultSyntheticLorentzian}
\end{figure}

The synthetic Lorentzian spectrum and the obtained area-weighted mean Lorentzian line width posterior distribution are illustrated in Figure \ref{im:resultSyntheticLorentzian}.
The Gaussian spectrum and the corresponding posterior distribution are shown in Figure \ref{im:resultSyntheticGaussian}.
In Figure \ref{im:resultSyntheticVoigt}, the Voigt spectrum and the posterior of the mean Lorentzian line width can be seen.
The posterior distribution for the Lorentzian spectrum in Figure \ref{im:resultSyntheticLorentzian} contains the true value for the area-weighted mean Lorentzian width.
The same is true for the Gaussian spectrum in Figure \ref{im:resultSyntheticGaussian} which exhibits also a far higher degree of uncertainty for the estimate.
The Voigt spectrum posterior in Figure \ref{im:resultSyntheticVoigt} likewise contains the true value of the mean width, with similar uncertainty to the Lorentzian spectrum.
The results for the Gaussian and Voigt spectrum are presented in Appendix \ref{sec:appendix}.

\hl{We present the results of the sensitivity analysis in Figure \ref{im:resultTruncationSimulation}.
The results show that the mean and 95\% confidence interval estimates converge and stay consistent with $P \geq 20$.
The confidence interval estimates experience blowup with truncation parameter $P < 20$.}
\begin{table}
\caption{Distributions of the spectrum parameters $( \bm{a}, \bm{l}, \bm\gamma, \bm\sigma)$ and measurement standard deviation $\sigma_\epsilon$ used to generate $M$ line shapes for generating the synthetic Lorentzian and Gaussian spectra. The top row corresponds to the prior distributions for the Lorentzian spectrum and the bottom row to the prior distributions for the Gaussian spectrum.}
\centering
\begin{tabular}{c c c c}
\toprule
Area & Location & Width & Noise level \\
\midrule
$a_m \sim \mathcal{U}( 1, 30)$ & $ l_m \sim \mathcal{U}( 1625, 1675) $ & $ \gamma_m \sim \mathcal{U}( 2.5, 20) $ & $ \sigma_\epsilon = 0.05 \times \max \bm{S} $\\
$a_m \sim \mathcal{U}( 1, 30)$ & $ l_m \sim \mathcal{U}( 1625, 1675) $ & $ \sigma_m \sim \mathcal{U}( 10, 30) $ & $ \sigma_\epsilon = 0.05 \times \max \bm{S} $\\
\bottomrule
\end{tabular}
\label{tb:spectrumParameters}
\end{table}
\section{Experimental results}
\label{sec:experimental}
In addition to the synthetic validation spectra, we apply the method to an experimental Raman spectrum of $\beta$-carotene \cite{Hsiung:2015}.
We found that the results were sensitive to the selected region of data used for the analysis.
Wide regions of data where the spectrum has multiple individual areas of peaks exhibit blowup for the posterior confidence intervals.
Smaller, more constrained intervals with fewer features present are better handled by the algorithm, producing worthwhile estimates for the mean Lorentzian line width and its confidence intervals.
We present \hl{posterior mean and confidence intervals} for different selections of data in Table \ref{tb:resultRegions}.
As a showcase, we illustrate the \hl{obtained posterior distribution} for a single selection of data in Figure \ref{im:resultExperimental}.
\hl{Posterior distribution illustrations} for the other selections of data are shown in \ref{sec:appendix}.
\begin{table}
\caption{Results obtained for different choices for the experimental $\beta$-carotene Raman spectrum for different regions of data.}
\centering
\begin{tabular}{c c c c}
\toprule
Interval [cm$^{-1}$] & Interval width [cm$^{-1}$] & Posterior mean & 95\% Confidence interval \\
\midrule
$[1240, 1060]$ & $180$ & $23.2$ & $[19.5, 27.0]$ \\
$[1300, 1060]$ & $240$ & $36.2$ & $[26.3, 46.1]$ \\
$[1360, 900]$ & $460$ & $61.5$ & $[44.2, 78.8]$ \\
$[1615, 900]$ & $715$ & $2.9 \cdot 10^6$ & $[0, 1.4 \cdot 10^7]$ \\
$[1615, 1060]$ & $555$ & $260.4$ & $[0, 3.8 \cdot 10^3]$ \\
$[1615, 1360]$ & $255$ & $24.1$ & $[22.3, 26.0]$ \\
\bottomrule
\end{tabular}
\label{tb:resultRegions}
\end{table}
\begin{figure}[H]
    \centering
    \includegraphics[width = \textwidth]{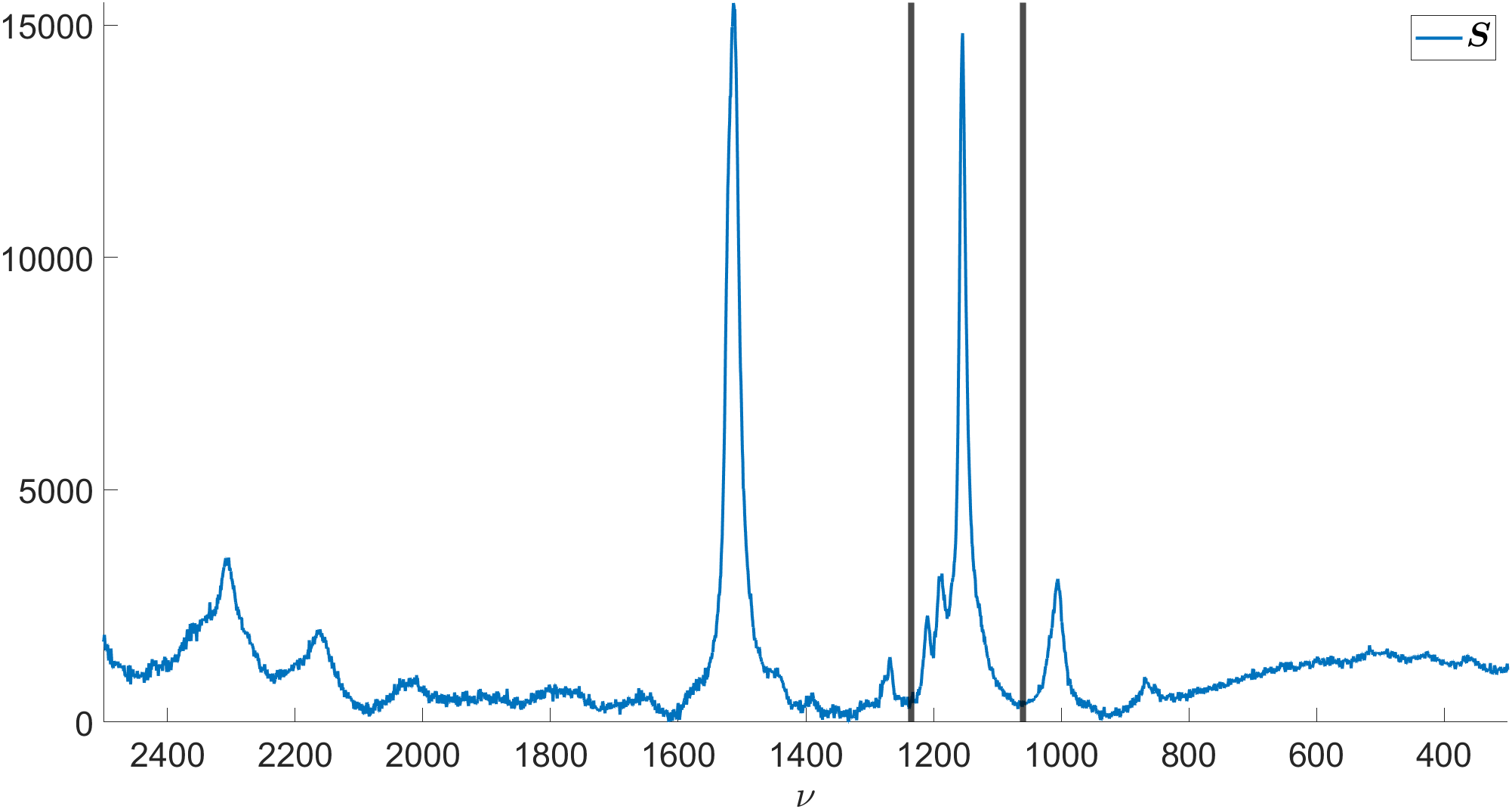}
    \caption{An experimental spectrum of $\beta$-carotene in blue. Black horizontal lines show the selected region ($[1240, 1060]$) of the spectrum.}
    \label{im:spectrumExperimental}
\end{figure}
\begin{figure}[H]
    \centering
    \includegraphics[width = \textwidth]{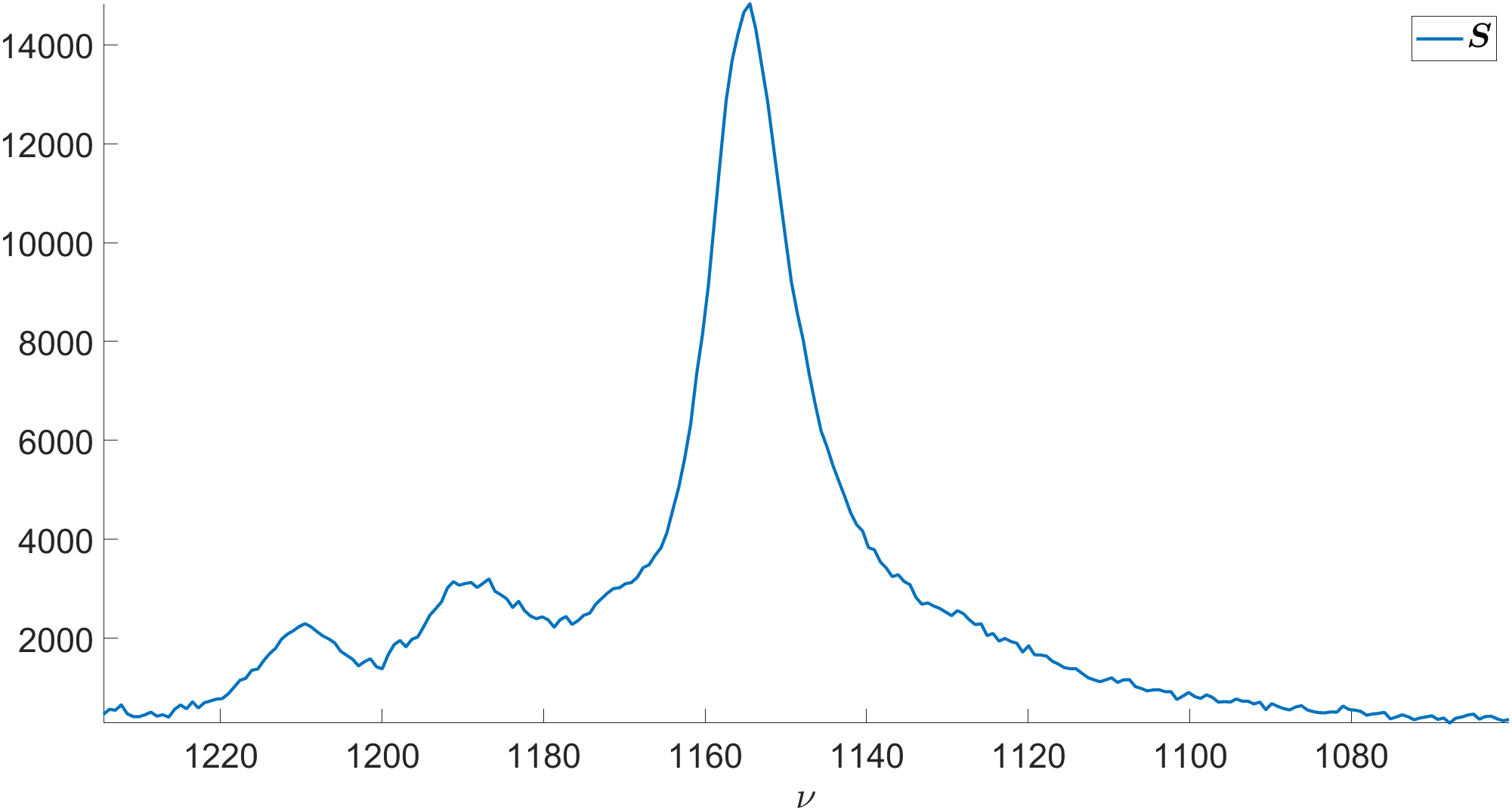}
    \includegraphics[width = \textwidth]{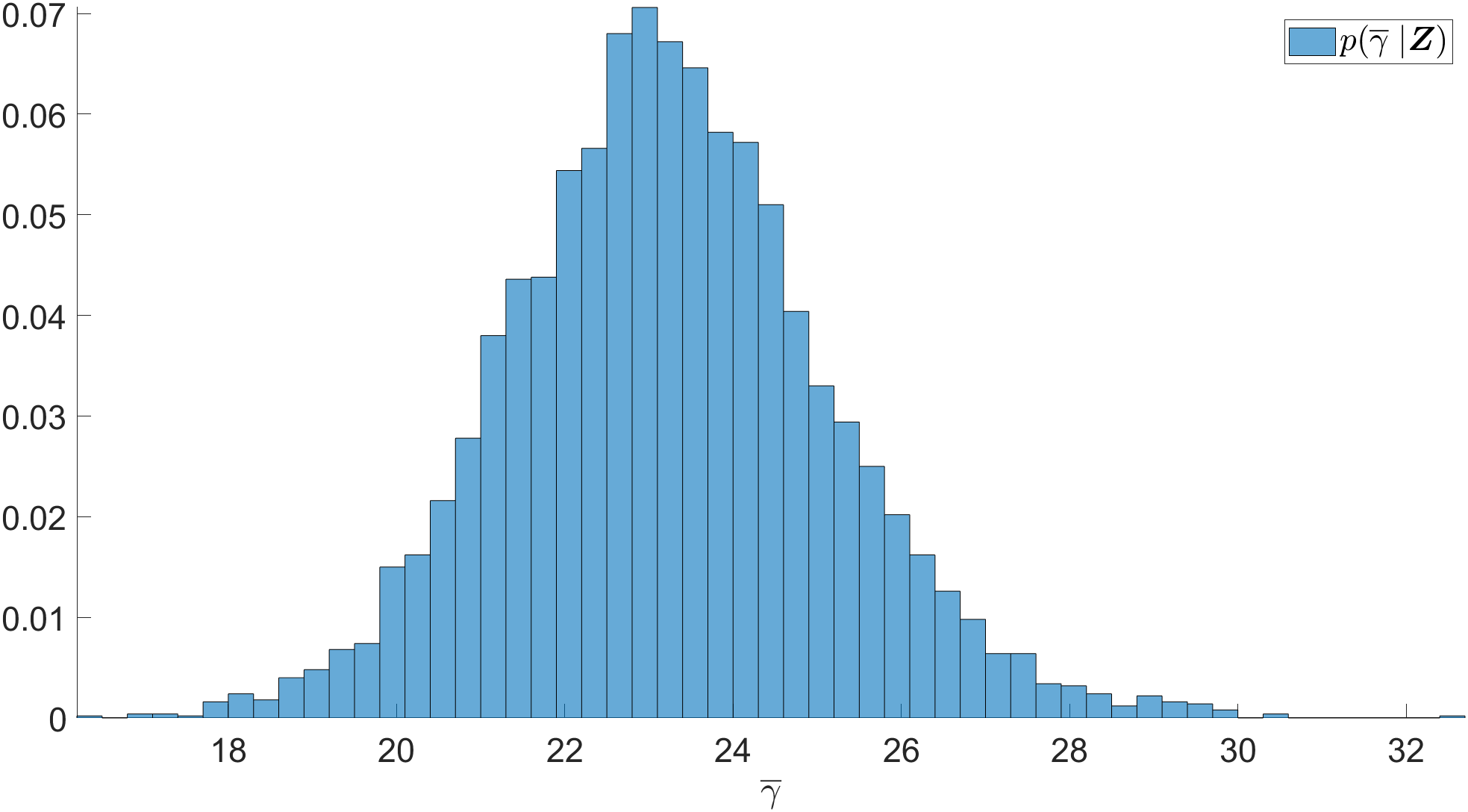}
    \caption{On top, the region between $[1240, 1060]$ of the experimental spectrum of $\beta$-carotene shown in Figure \ref{im:spectrumExperimental}. On bottom, the obtained posterior distribution $p( \overline{\gamma} \mid \bm{Z} )$ for the area-weighted mean Lorentzian line width.}
    \label{im:resultExperimental}
\end{figure}
\section{Conclusions}
\label{sec:conclusions}
We have proposed and implemented a statistical algorithm for estimating the area-weighted mean Lorentzian line width in spectra consisting of Lorentzian, Gaussian, or Voigt line shapes.
This extends previously introduced methodology by automating the estimation process and by enabling uncertainty quantification with Gaussian processes and Markov chain Monte Carlo methods in a two-stage implementation.
The method is validated with three synthetic spectra.
In all three cases, the proposed method yields posterior distributions that contain the true parameter value used to generate the synthetic spectra.
Additionally, the method is applied to an experimental \hl{Raman} spectrum of $\beta$-carotene.
\hl{We anticipate that our method would be broadly applicable to many different types of electromagnetic spectra, including UV-Vis \cite{Antonov:2000}, coherent anti-Stokes Raman scattering \cite{Harkonen:2020}, or X-ray spectroscopy \cite{suuronen2020enhancing}.}

The method allows for some straightforward modifications.
The squared exponential covariance function in Eq.~\eqref{eq:squaredExpKernel} can be replaced with the more general Matérn covariance function with a fixed regularity parameter $\nu$ or so that the regularity parameter is also estimated with MCMC.
The squared exponential covariance function is a special case of the Matérn covariance function when $\nu \rightarrow \infty$.
Discussion on different covariance functions can be found for example in \cite{Rasmussen2005}.
For computational speed-up, MCMC sampling can be replaced with optimization.
This would yield a single point estimate for the GP parameters.
This was however found to have an effect on the resulting mean Lorentzian line width posterior distribution, $ p( \overline{\gamma} \mid \bm{Z} )$.
An example of a more involved extension would be using mixtures of Gaussian process experts where multiple Gaussian processes are used to model different parts of data with individual Gaussian processes \cite{Volker:2000, Zhang:2019, Harkonen:2022}.
This could come into play with spectra with wide tails or where the spectrum is not adequately modelled by a single GP.

The statistical approach also enables alternative approaches to further validate the approach with simulation-based calibration \cite{ Talts:2020, McLeod:2021}.
This has been previously considered in the context of spectrum analysis by the authors \cite{Harkonen:2023}.
\section*{Acknowledgements}
This work was supported by the Research Council of Finland through the Flagship of Advanced Mathematics for Sensing, Imaging and Modelling (decision number 359183).
\bibliography{ref.bib}
\newpage
\appendix
\section{Additional results}
\label{sec:appendix}
\begin{figure}[H]
    \centering
    \includegraphics[width = \textwidth]{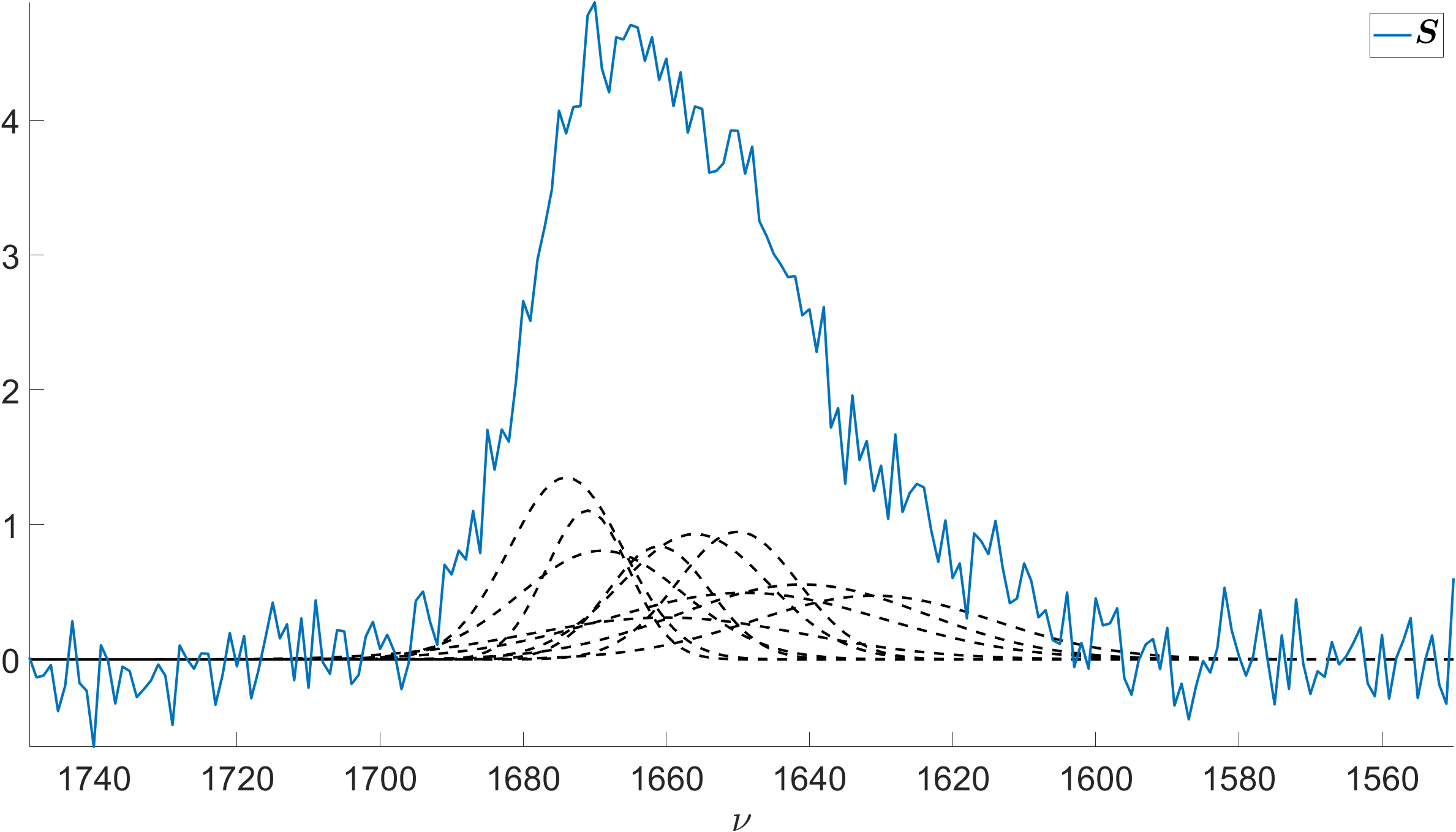}
    \includegraphics[width = \textwidth]{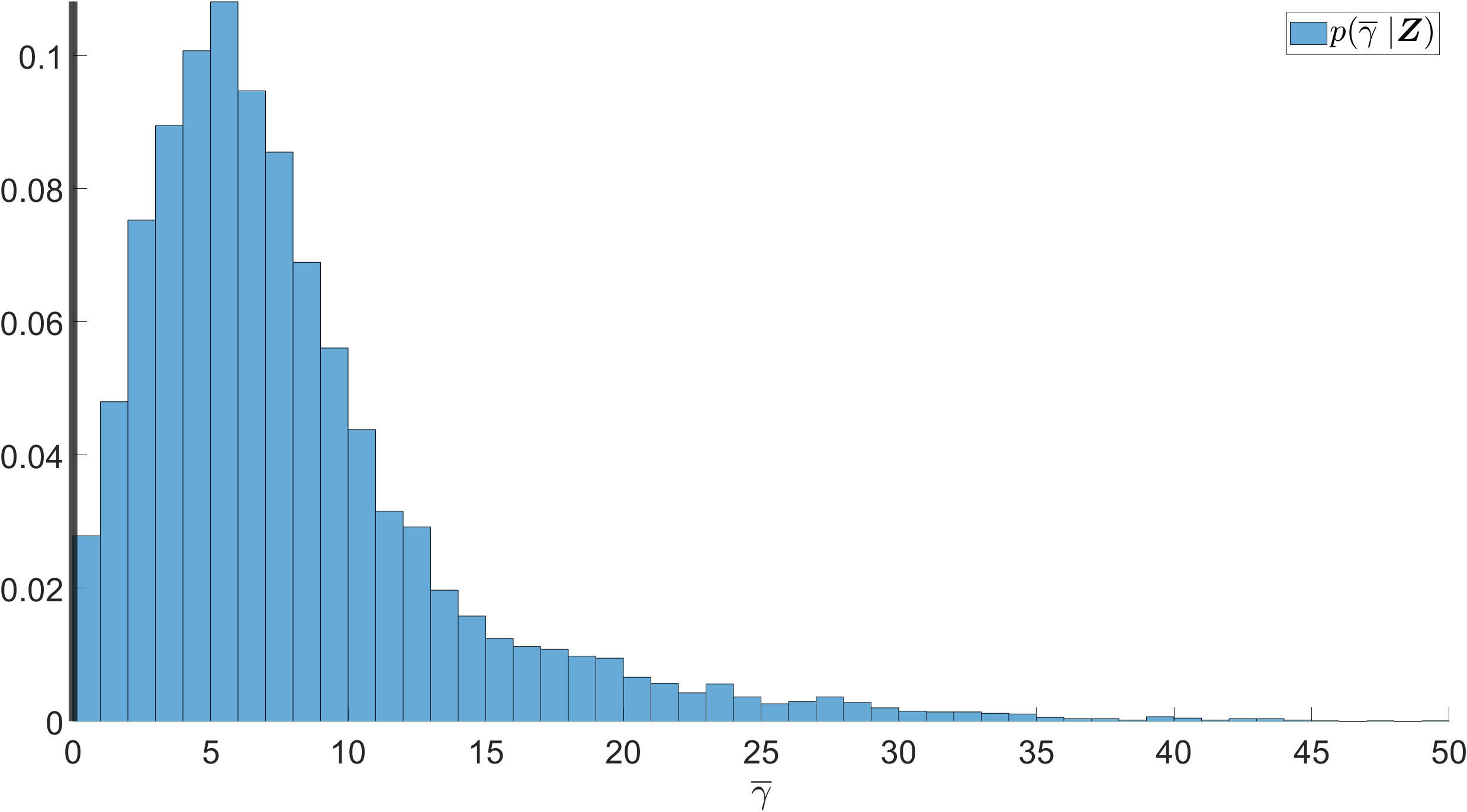}
    \caption{On top, a spectrum consisting of $ M = 10 $ Gaussian line shapes in blue along with the individual overlapping line shapes illustrated with black dashed lines. On bottom, the obtained posterior distribution $p( \overline{\gamma} \mid \bm{Z} )$ for the area-weighted mean Lorentzian line width. The true value for $\overline{\gamma} = 0$ used to generate the data is shown with a vertical solid line.}
    \label{im:resultSyntheticGaussian}
\end{figure}
\begin{figure}
    \centering
    \includegraphics[width = \textwidth]{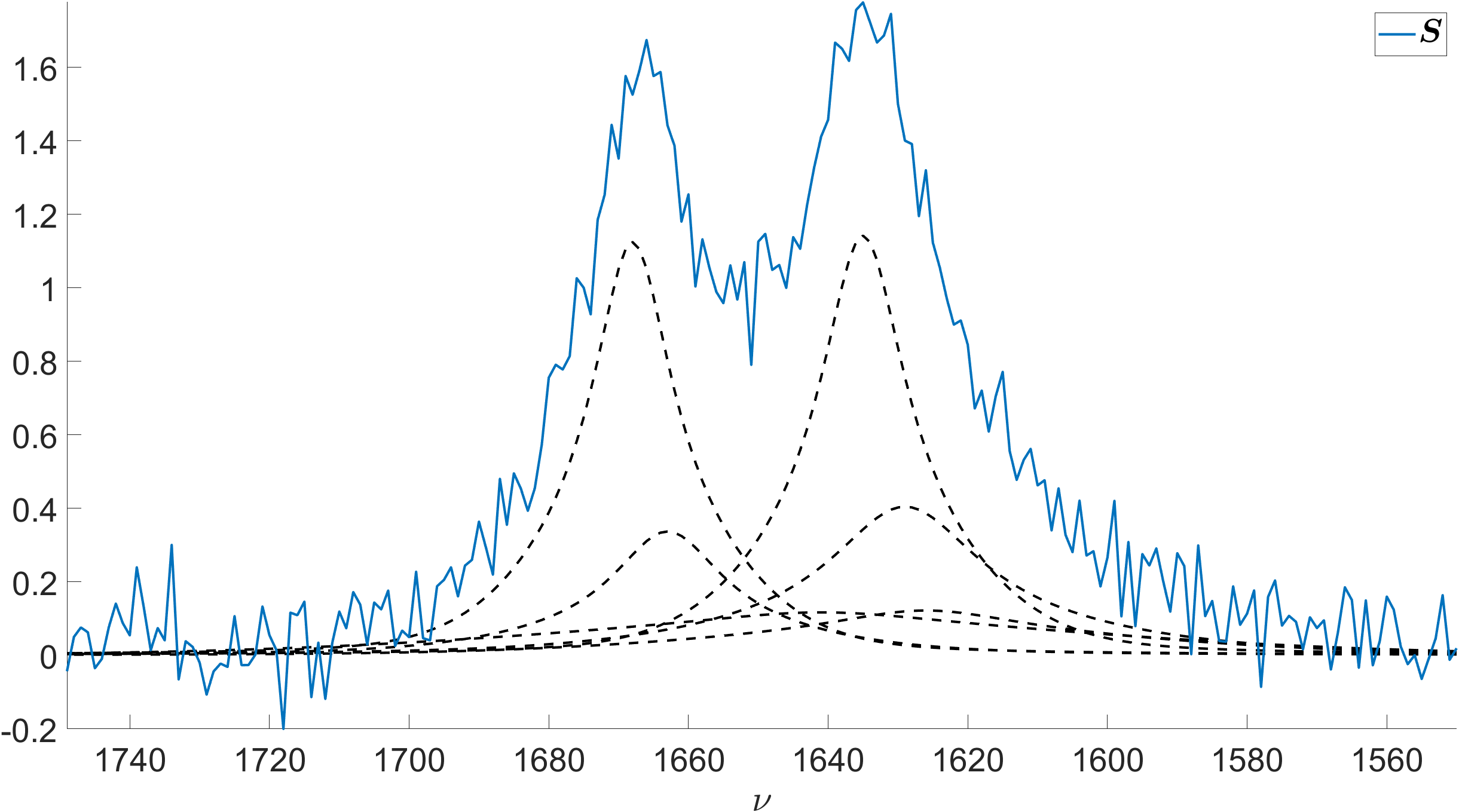}
    \includegraphics[width = \textwidth]{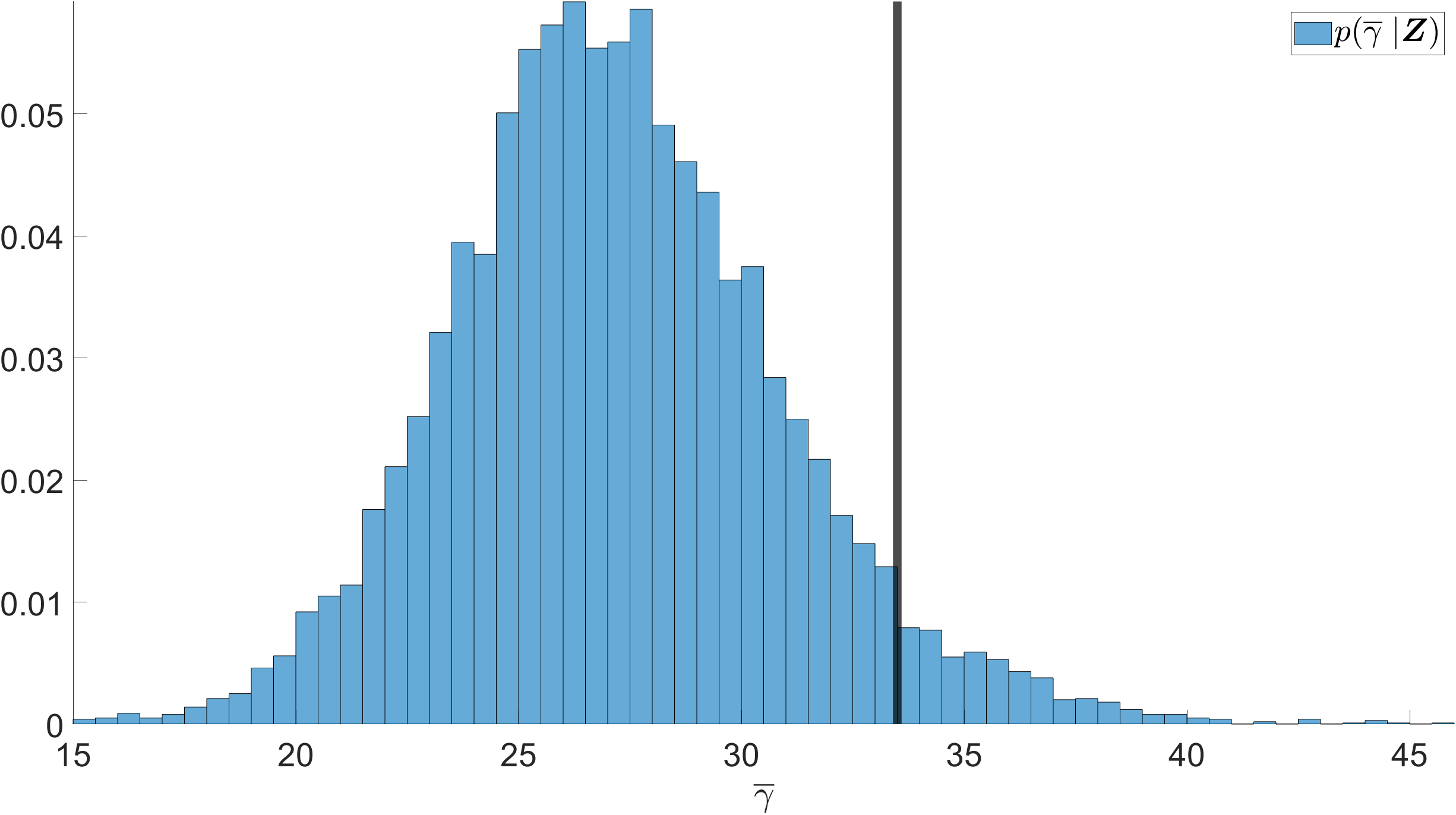}    \caption{On top, a spectrum consisting of $ M = 6 $ Voigt line shapes in blue along with the individual overlapping line shapes illustrated with black dashed lines. On bottom, the obtained posterior distribution $p( \overline{\gamma} \mid \bm{Z} )$ for the area-weighted mean Lorentzian line width. The true value for $\overline{\gamma} = 33.50$ used to generate the data is shown with a vertical solid line.}
    \label{im:resultSyntheticVoigt}
\end{figure}
\begin{figure}
    \centering
    \includegraphics[width = .9\textwidth]{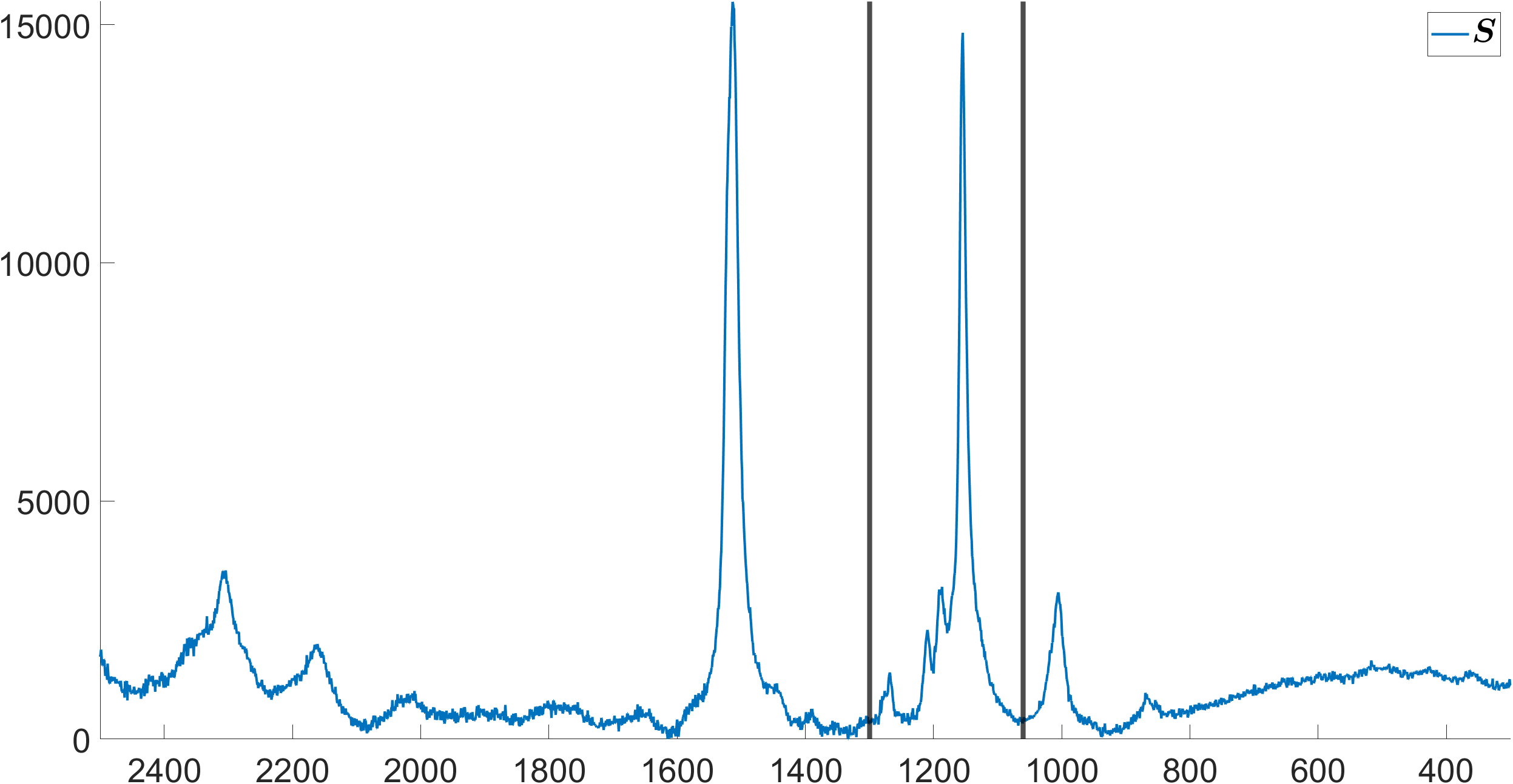}
    \includegraphics[width = .9\textwidth]{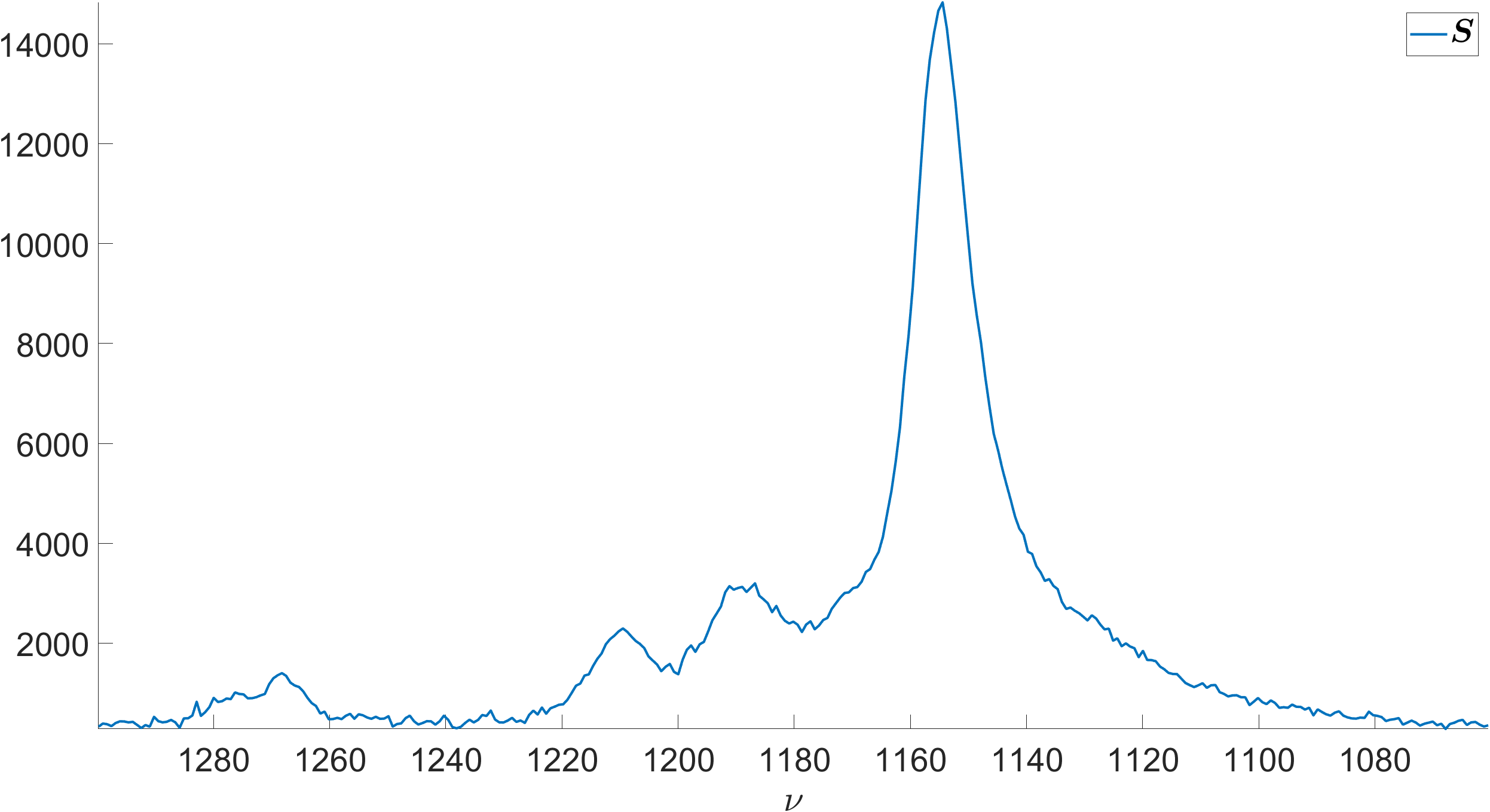}
    \includegraphics[width = .9\textwidth]{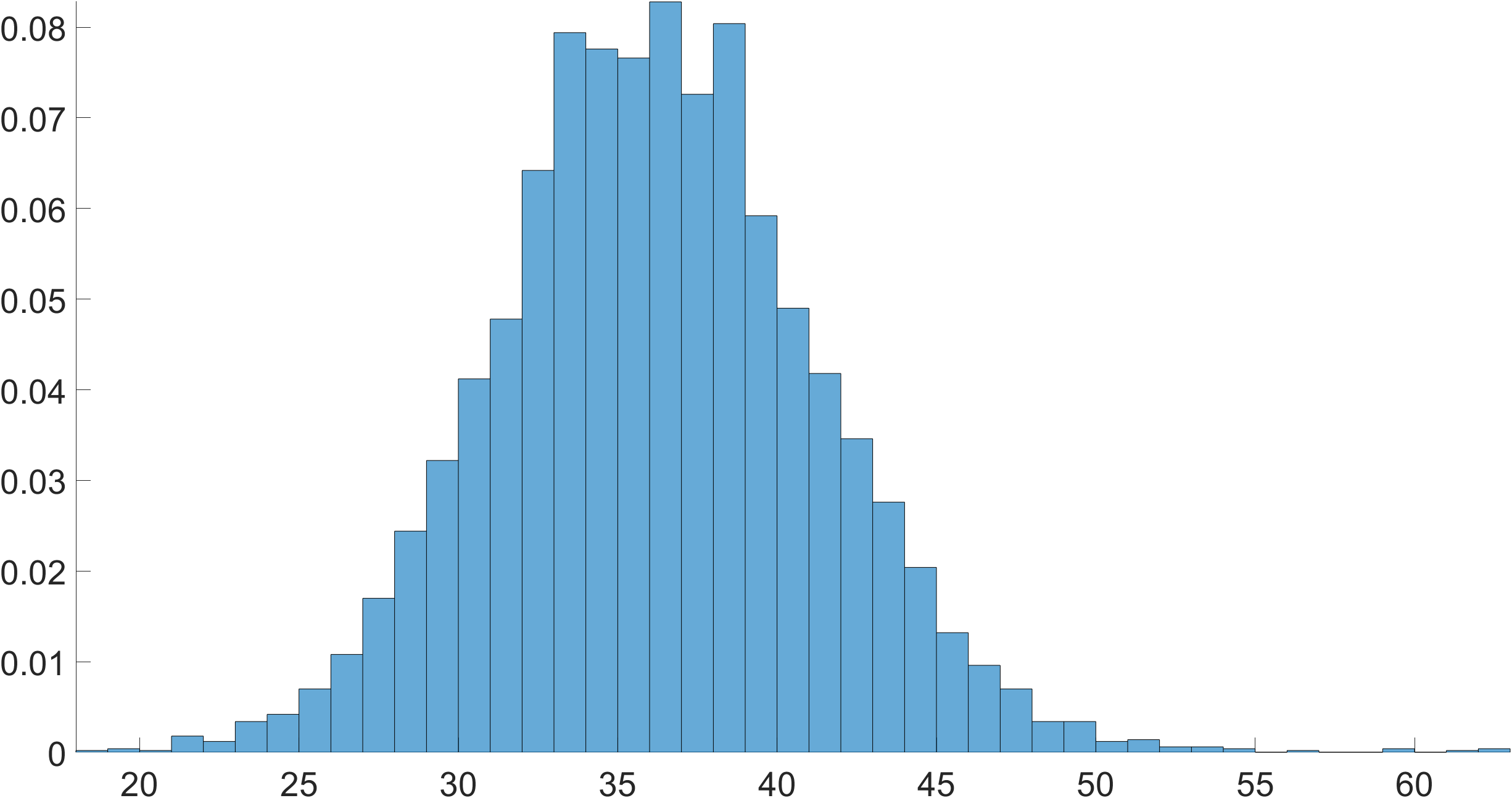}
    \caption{On top, an experimental spectrum of $\beta$-carotene with selection $[1300-1060]$ in the middle. On bottom, the obtained posterior distribution $p( \overline{\gamma} \mid \bm{Z} )$ for the area-weighted mean Lorentzian line width.}
    \label{im:resultExperimentalA1}
\end{figure}
\begin{figure}
    \centering
    \includegraphics[width = .9\textwidth]{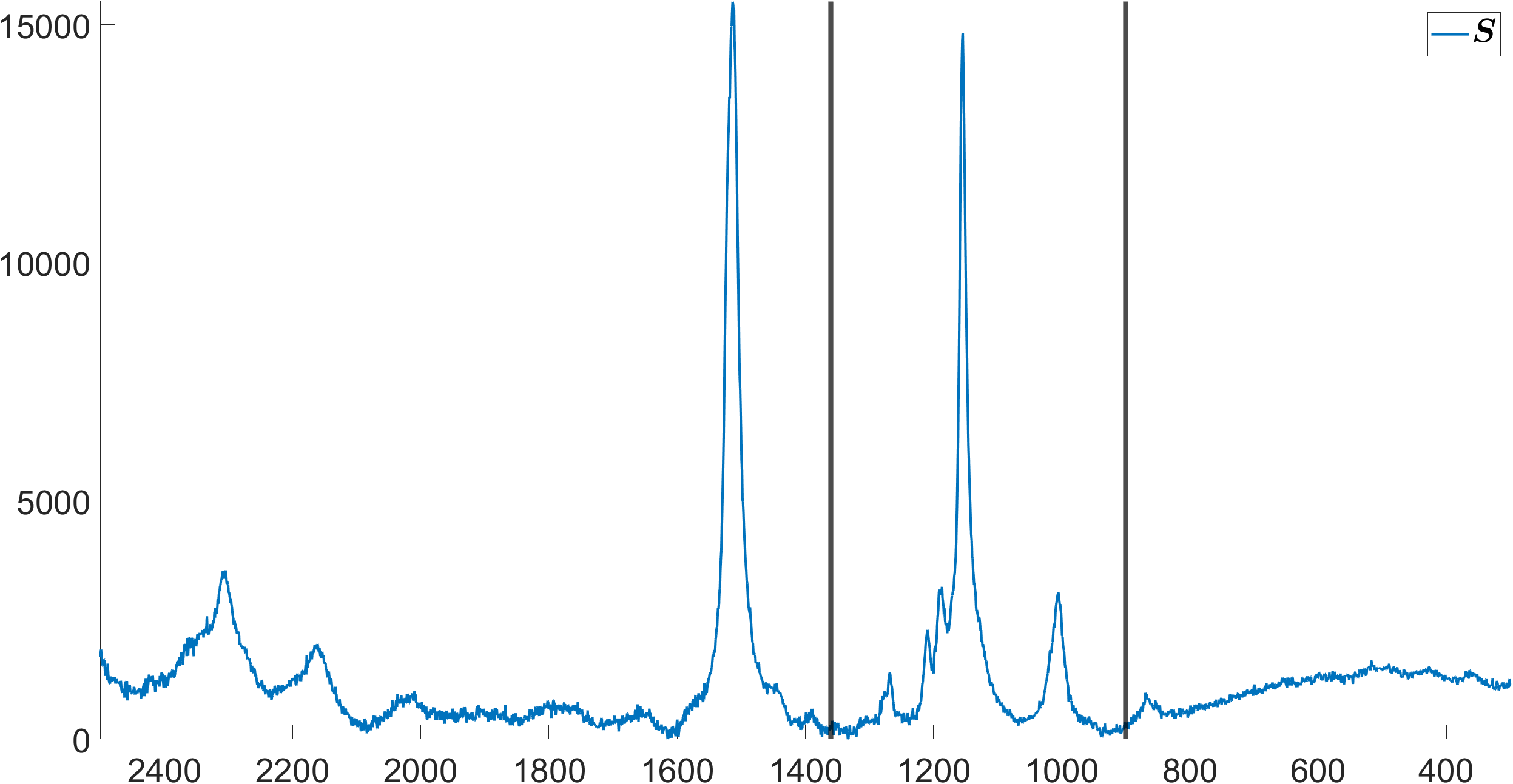}
    \includegraphics[width = .9\textwidth]{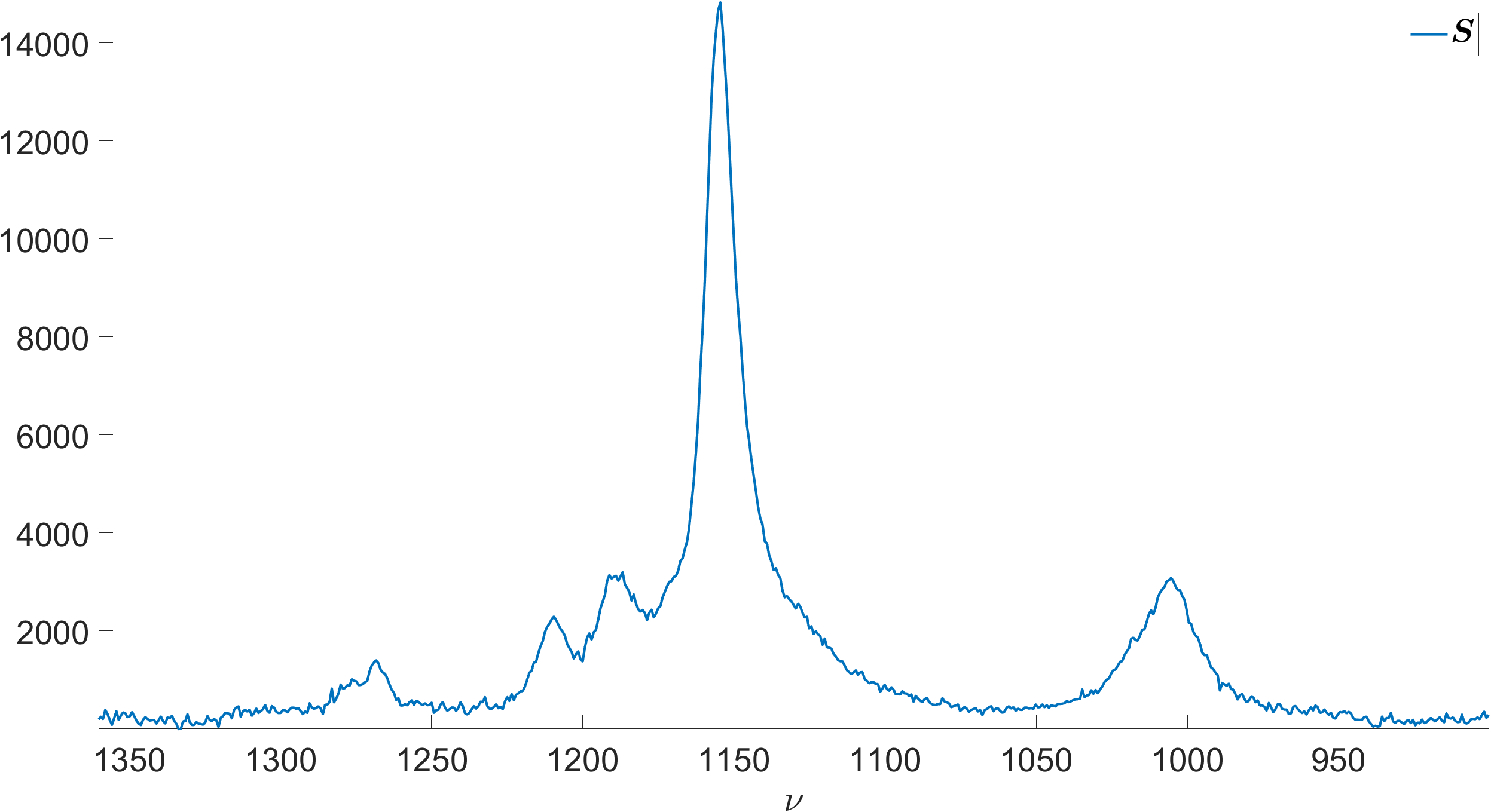}
    \includegraphics[width = .9\textwidth]{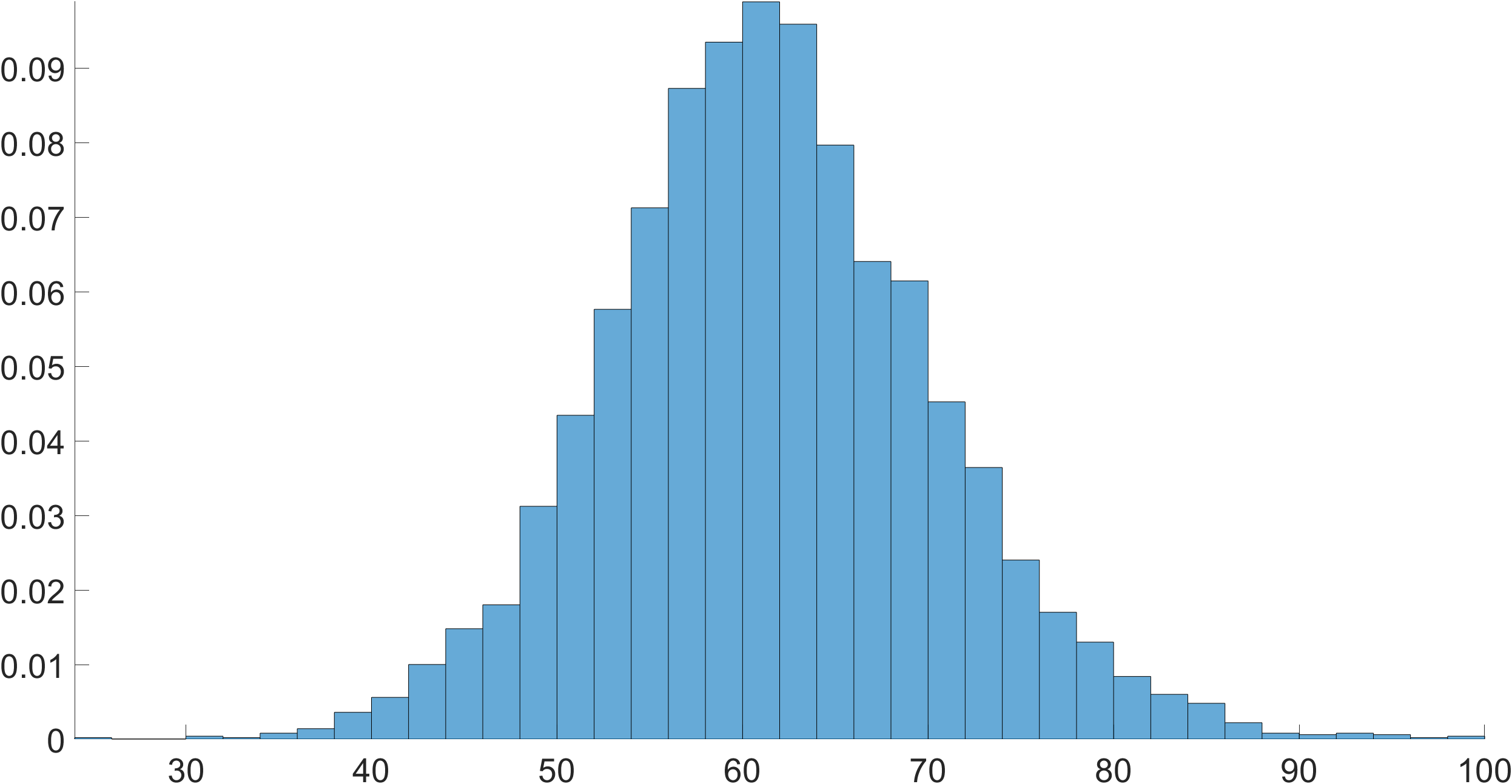}
    \caption{On top, an experimental spectrum of $\beta$-carotene with selection $[1360-900]$ in the middle. On bottom, the obtained posterior distribution $p( \overline{\gamma} \mid \bm{Z} )$ for the area-weighted mean Lorentzian line width.}
    \label{im:resultExperimentalA2}
\end{figure}
\begin{figure}
    \centering
    \includegraphics[width = .9\textwidth]{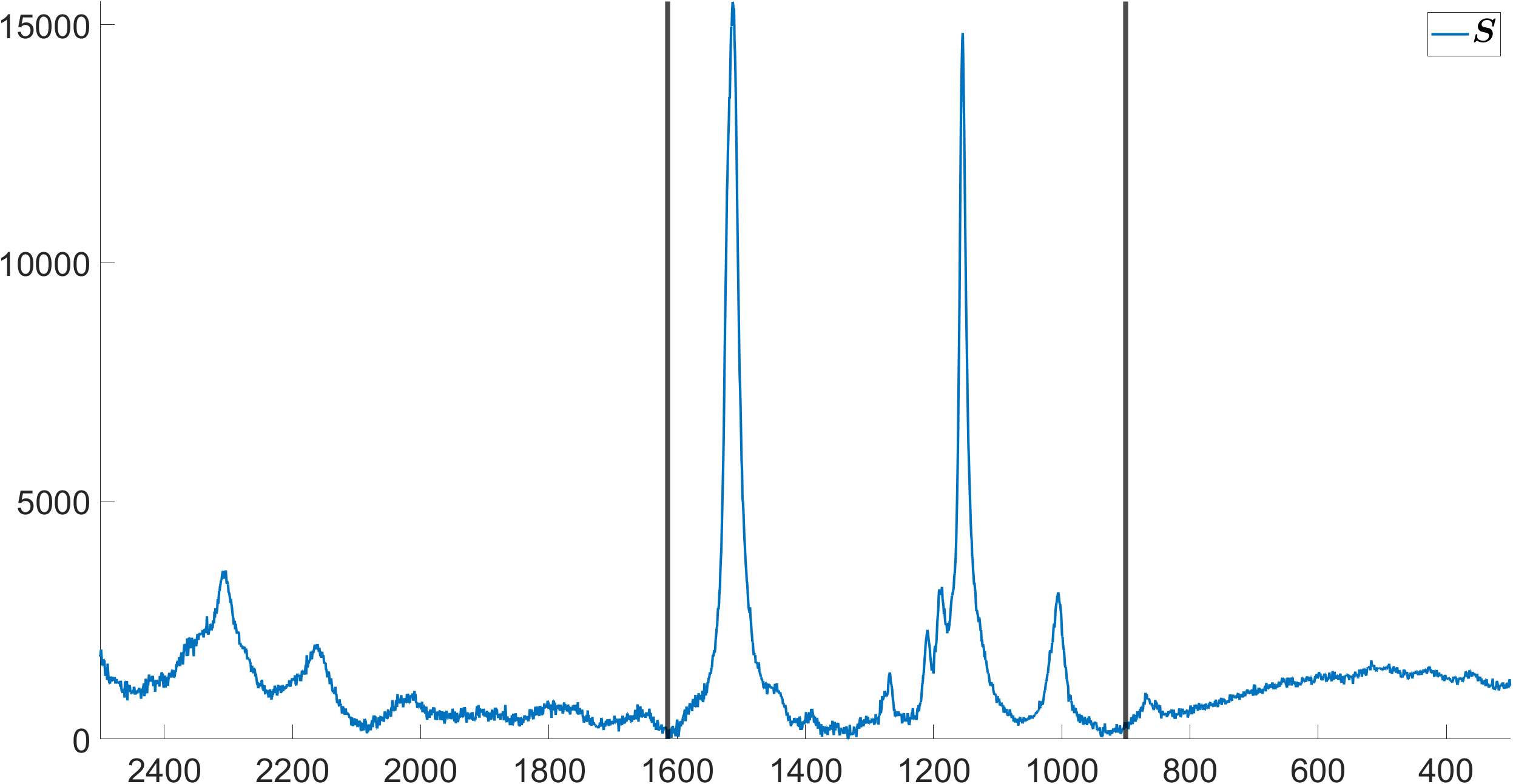}
    \includegraphics[width = .9\textwidth]{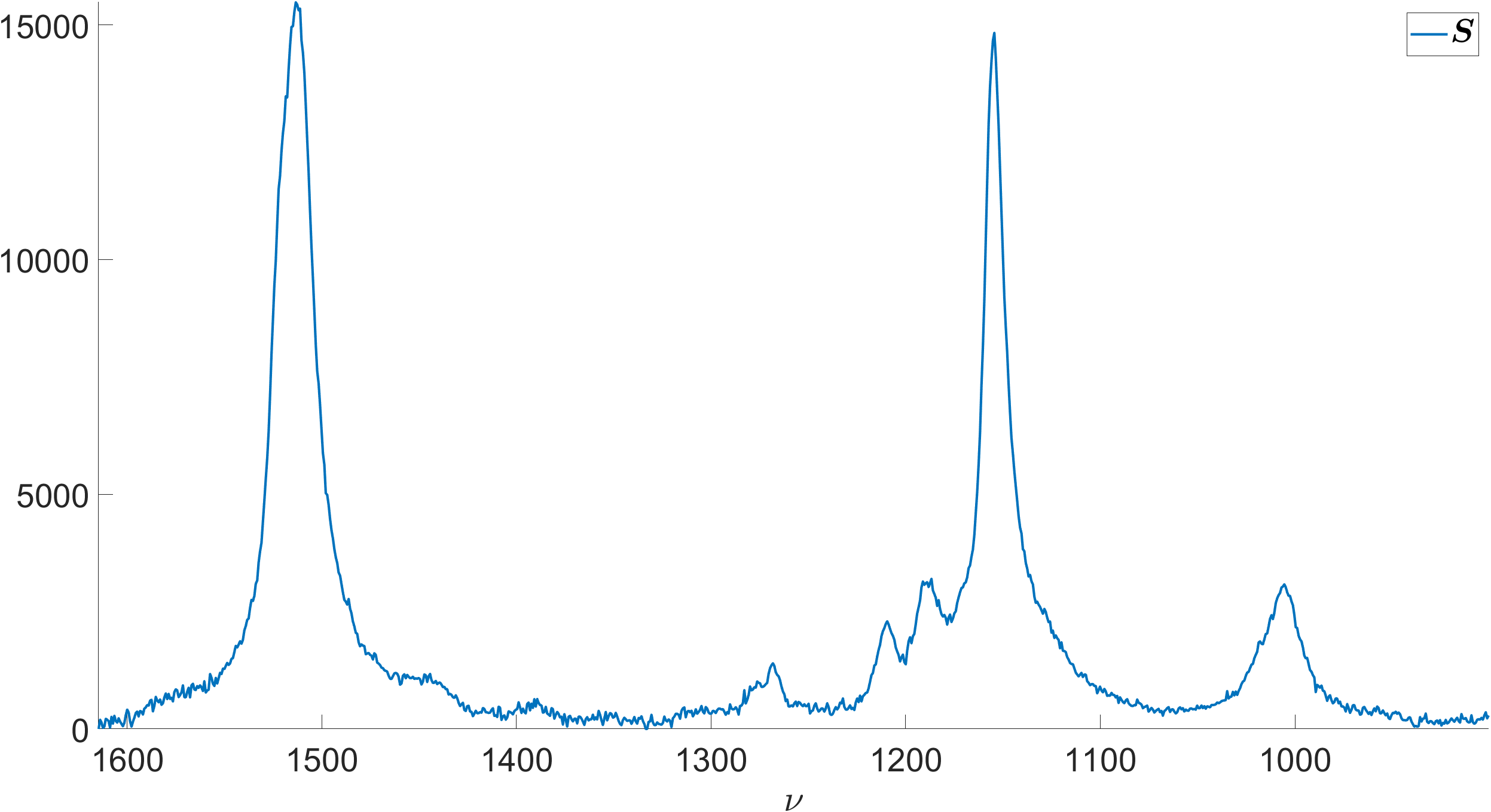}
    \includegraphics[width = .9\textwidth]{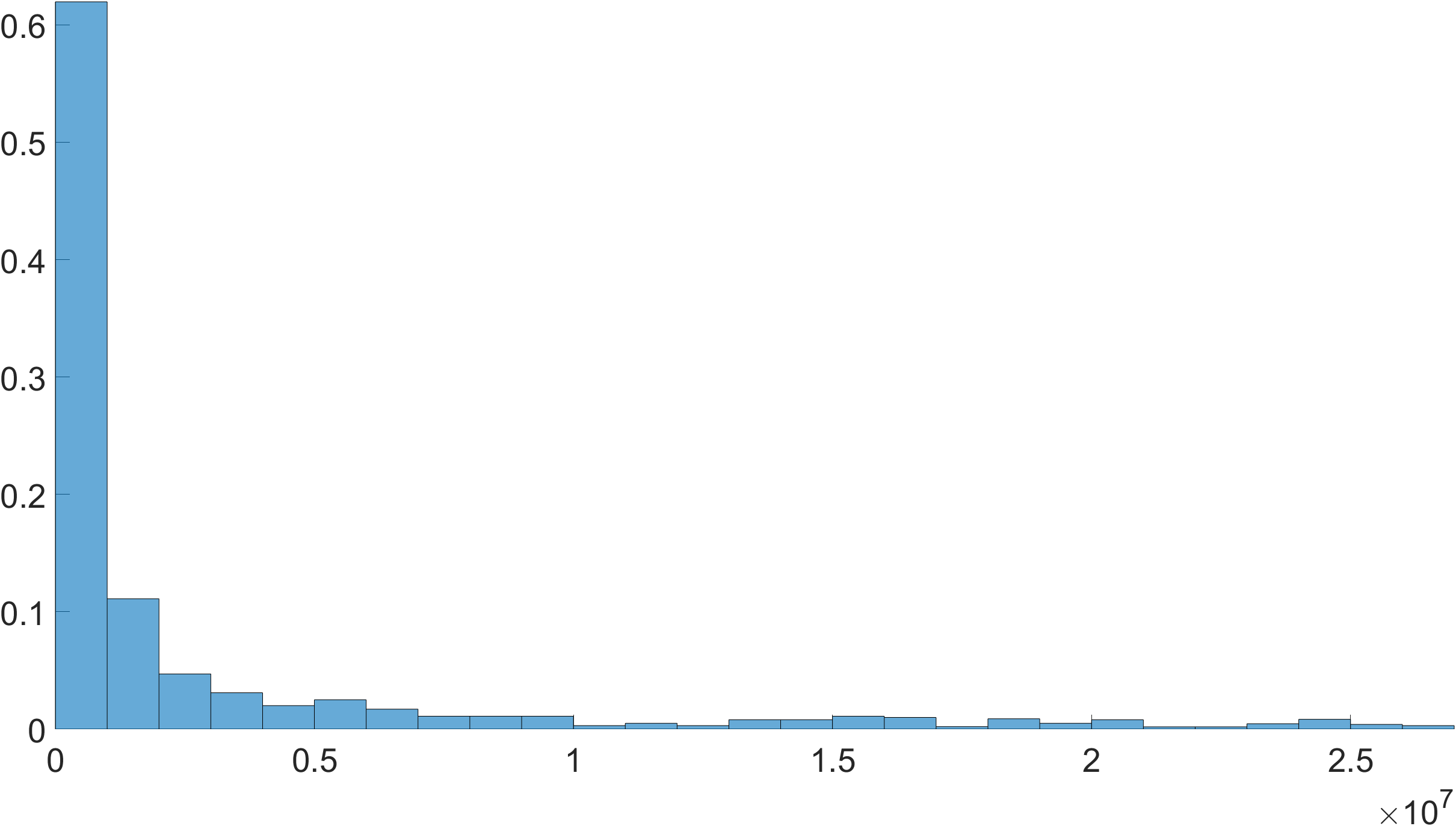}
    \caption{On top, an experimental spectrum of $\beta$-carotene with selection $[1615-900]$ in the middle. On bottom, the obtained posterior distribution $p( \overline{\gamma} \mid \bm{Z} )$ for the area-weighted mean Lorentzian line width.}
    \label{im:resultExperimentalA3}
\end{figure}
\begin{figure}
    \centering
    \includegraphics[width = .9\textwidth]{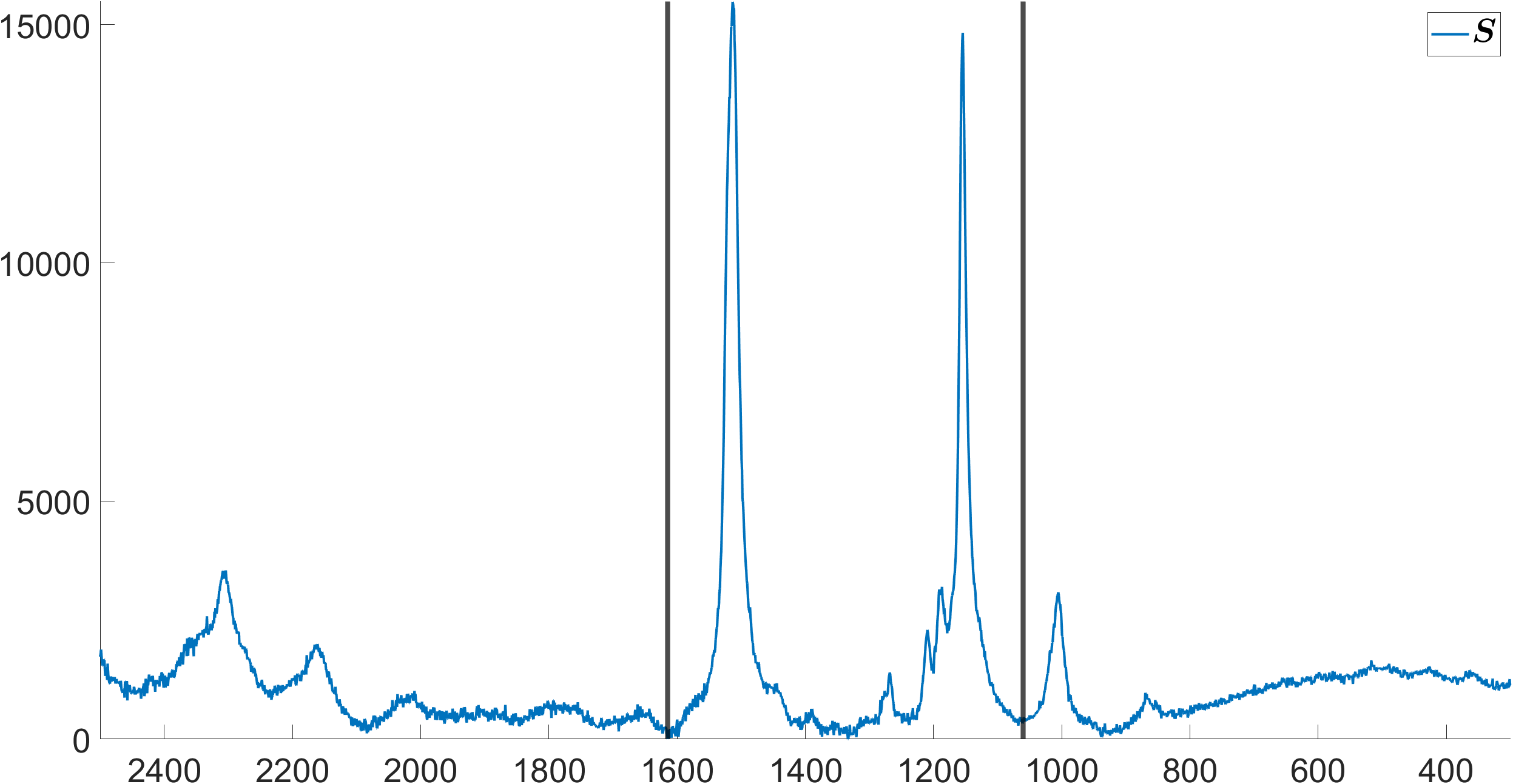}
    \includegraphics[width = .9\textwidth]{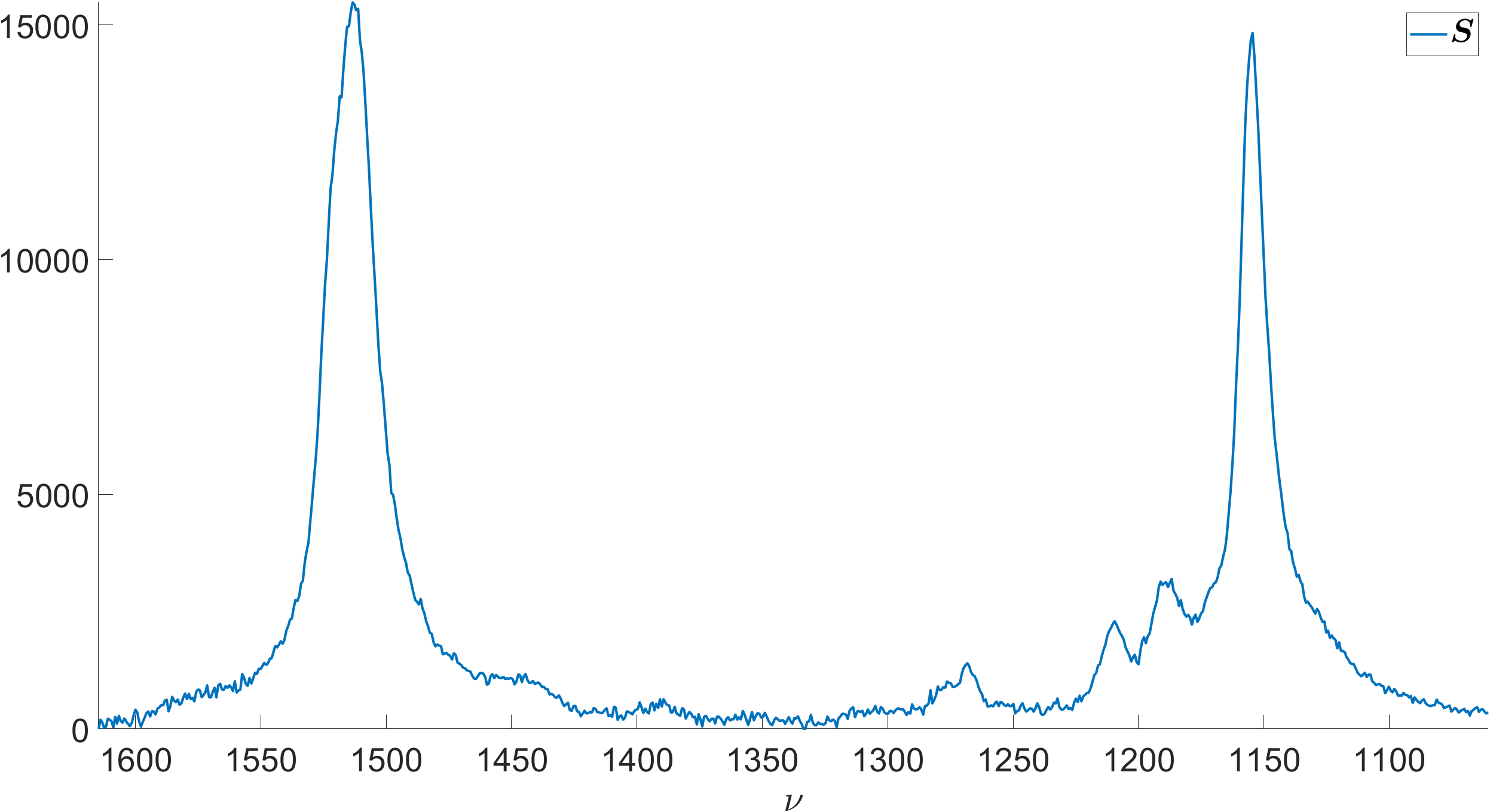}
    \includegraphics[width = .9\textwidth]{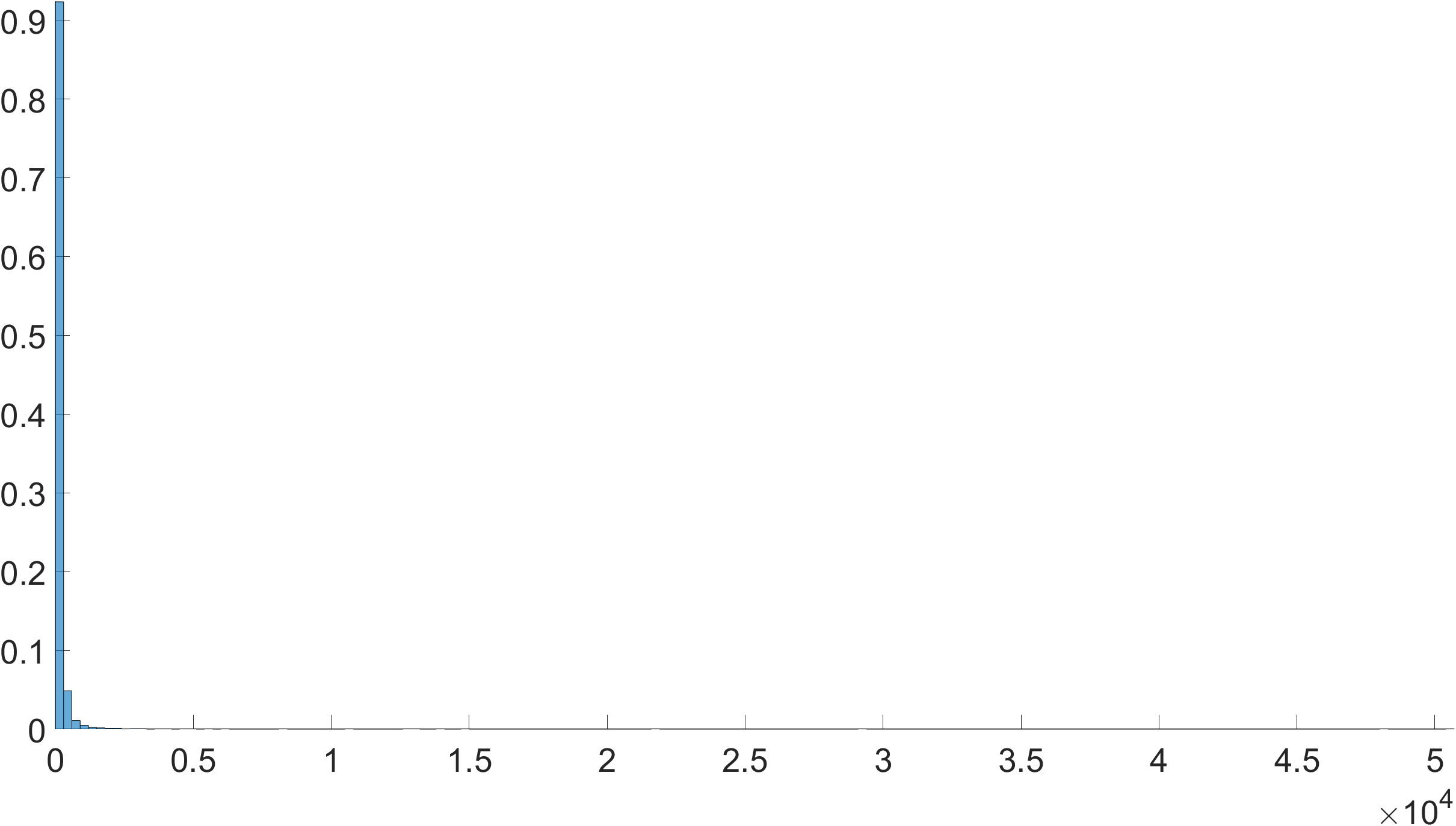}
    \caption{On top, an experimental spectrum of $\beta$-carotene with selection $[1615-900]$ in the middle. On bottom, the obtained posterior distribution $p( \overline{\gamma} \mid \bm{Z} )$ for the area-weighted mean Lorentzian line width.}
    \label{im:resultExperimentalA4}
\end{figure}
\begin{figure}
    \centering
    \includegraphics[width = .9\textwidth]{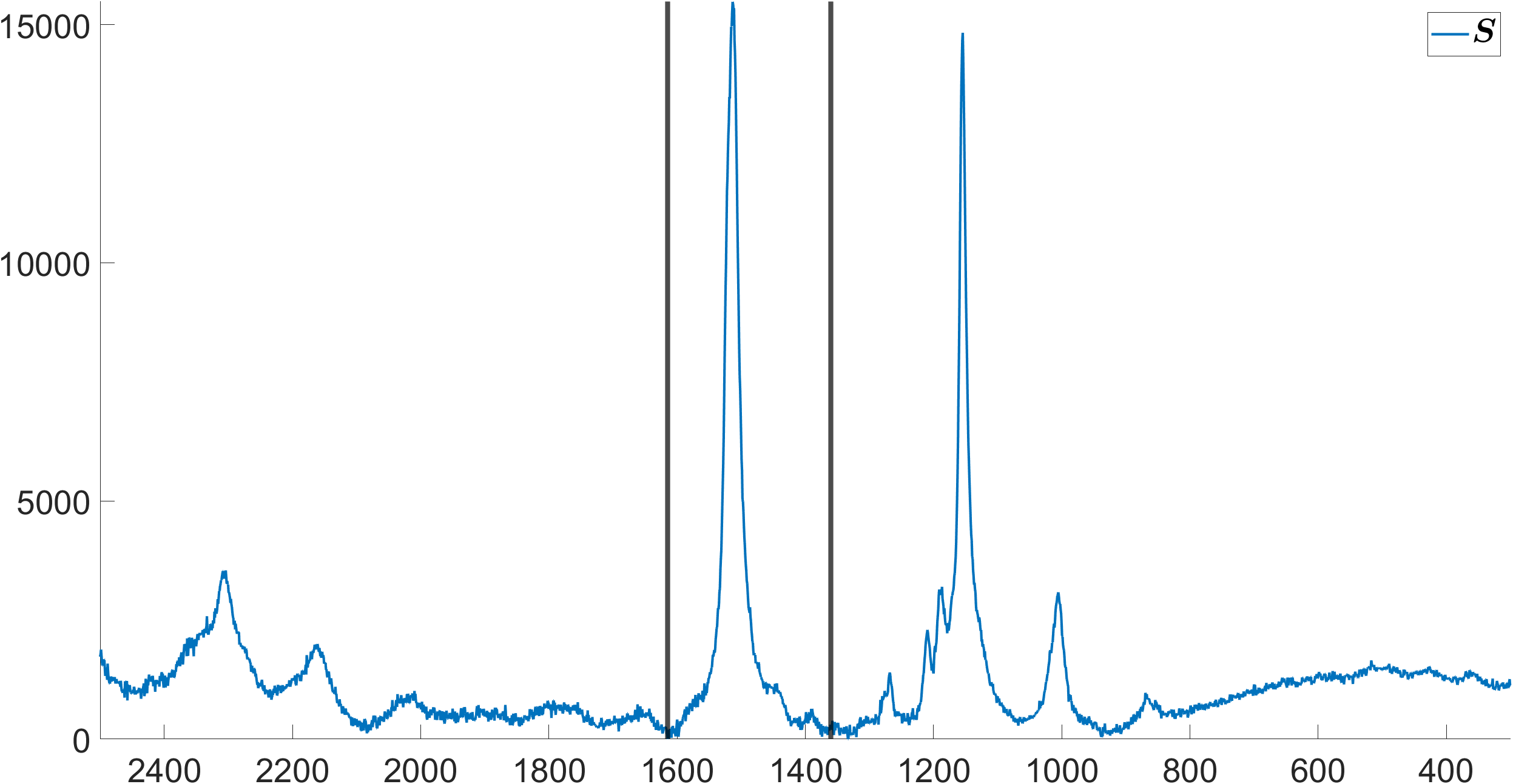}
    \includegraphics[width = .9\textwidth]{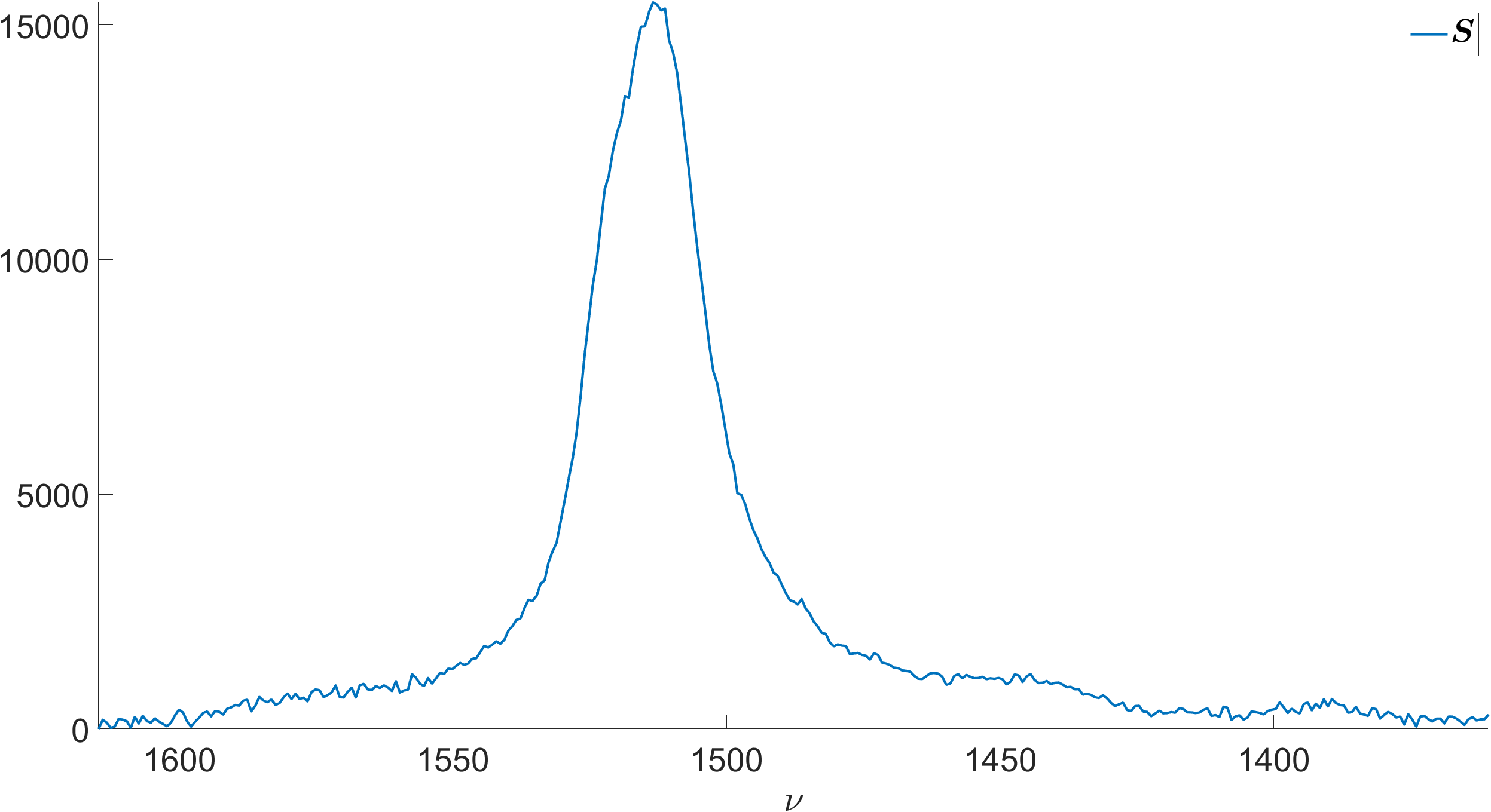}
    \includegraphics[width = .9\textwidth]{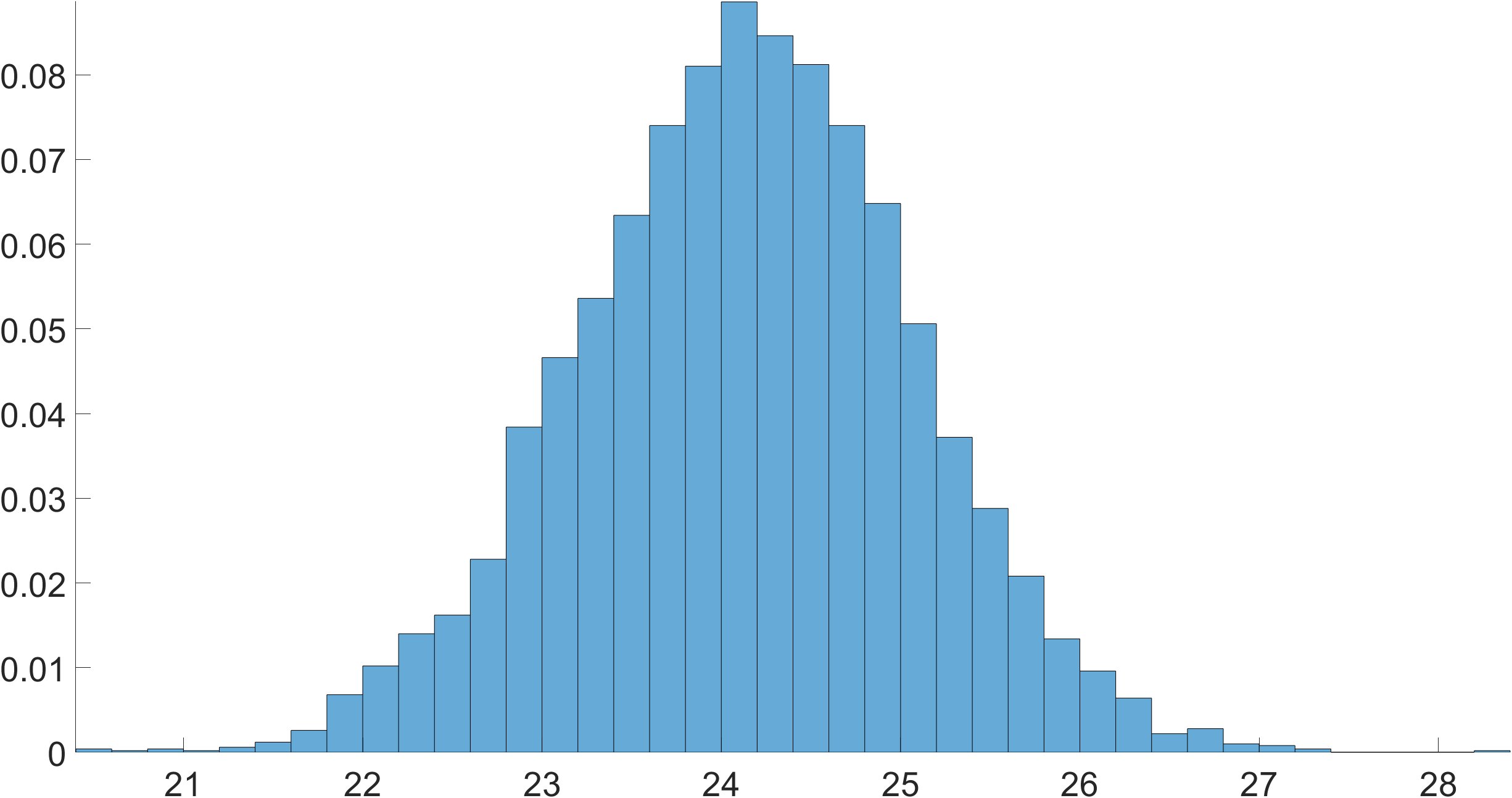}
    \caption{On top, an experimental spectrum of $\beta$-carotene with selection $[1615-900]$ in the middle. On bottom, the obtained posterior distribution $p( \overline{\gamma} \mid \bm{Z} )$ for the area-weighted mean Lorentzian line width.}
    \label{im:resultExperimentalA5}
\end{figure}
\begin{figure}
    \centering
    \includegraphics[width = .9\textwidth]{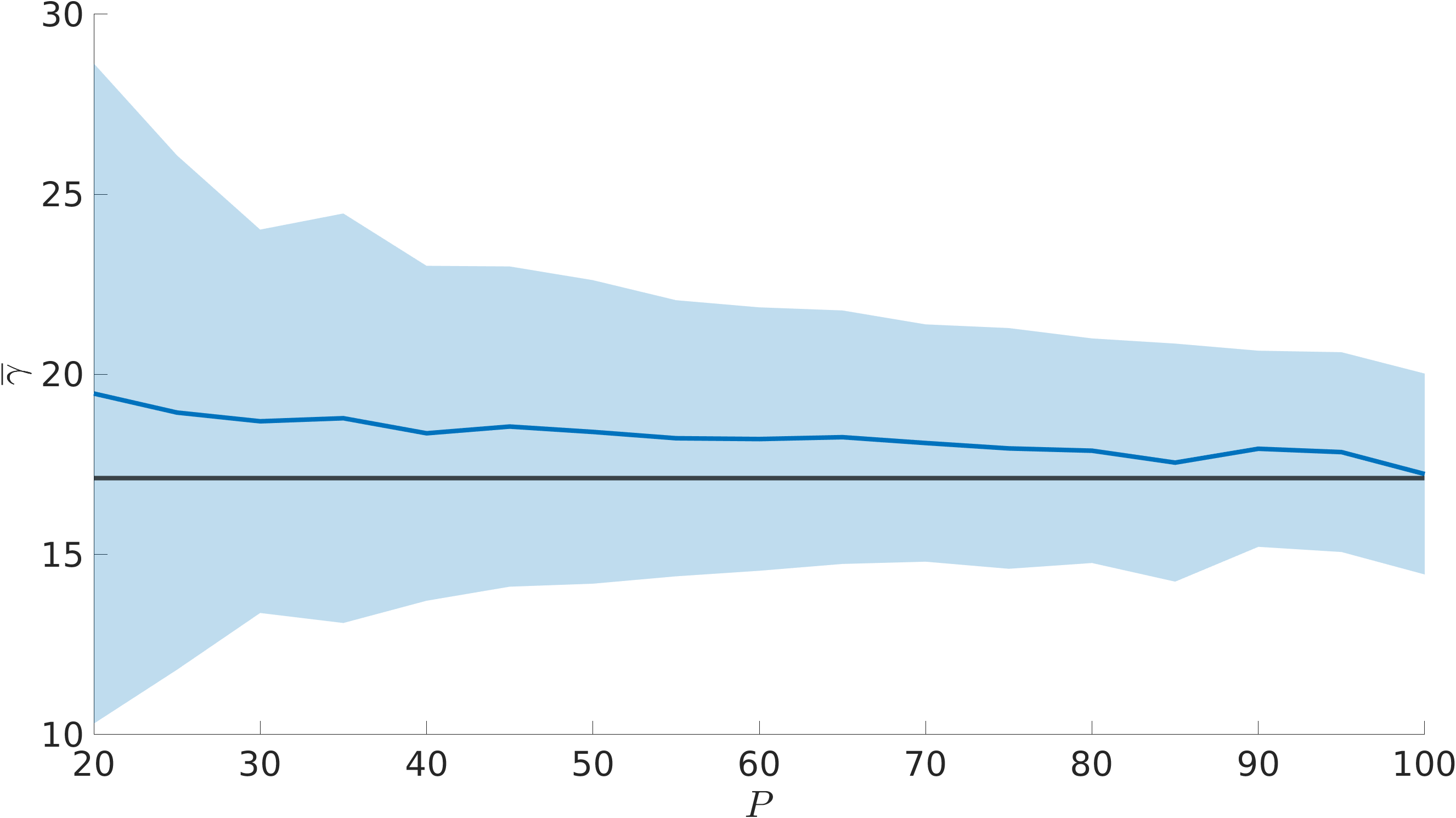}
    \caption{\hl{Posterior mean (solid line) and 95\% confidence intervals (shaded blue region) for the mean Lorentzian line width $\overline{\gamma}$ for truncation parameter $P \in ( 20, 25, \dots, 95, 100)^T$ for the synthetic spectrum consisting of $M = 8$ Lorentzian line shapes in Figure \ref{im:resultSyntheticLorentzian}. The true value for $\overline{\gamma} = 17.12$ used to generate the data is shown with a horizontal solid line. The mean and 95\% confidence intervals experience blowup with $P < 20$. The mean and confidence interval estimates converge and stay consistent with $P \geq 20$.}}
    \label{im:resultTruncationSimulation}
\end{figure}
\end{document}